\documentclass[twocolumn, showpacs, longbibliography, prl,aps, amsmath, amssymb,superscriptaddress]{revtex4-1}
\usepackage{amsmath}  
\usepackage{amsfonts} 
\usepackage{graphicx} 
\usepackage[colorlinks,linkcolor=blue,citecolor=blue,urlcolor=blue]{hyperref}

\begin{document}

\title{Probing a dissipative phase transition via dynamical optical hysteresis }

\author{S.R.K. Rodriguez} \email{said.rodriguez@lpn.cnrs.fr}
\affiliation {Centre de Nanosciences et de Nanotechnologies, CNRS, Univ. Paris-Sud, Universit\'{e} Paris-Saclay, C2N---Marcoussis, 91460 Marcoussis, France}

\author{W. Casteels}
\affiliation {Laboratoire Mat\'eriaux et Ph\'enom\`enes Quantiques,
Universit\'{e} Paris Diderot, Sorbonne Paris Cit\'{e} and CNRS, UMR
7162,  75205 Paris Cedex 13, France}

\author{F. Storme}
\affiliation {Laboratoire Mat\'eriaux et Ph\'enom\`enes Quantiques,
Universit\'{e} Paris Diderot, Sorbonne Paris Cit\'{e} and CNRS, UMR
7162,  75205 Paris Cedex 13, France}

\author{N. Carlon Zambon}
\affiliation {Centre de Nanosciences et de Nanotechnologies, CNRS, Univ. Paris-Sud, Universit\'{e} Paris-Saclay, C2N---Marcoussis, 91460 Marcoussis, France}

\author      {I. Sagnes}
\affiliation {Centre de Nanosciences et de Nanotechnologies, CNRS, Univ. Paris-Sud, Universit\'{e} Paris-Saclay, C2N---Marcoussis, 91460 Marcoussis, France}

\author      {L. Le Gratiet}
\affiliation {Centre de Nanosciences et de Nanotechnologies, CNRS, Univ. Paris-Sud, Universit\'{e} Paris-Saclay, C2N---Marcoussis, 91460 Marcoussis, France}

\author      {E. Galopin}
\affiliation {Centre de Nanosciences et de Nanotechnologies, CNRS, Univ. Paris-Sud, Universit\'{e} Paris-Saclay, C2N---Marcoussis, 91460 Marcoussis, France}

\author      {A. Lema\^{i}tre}
\affiliation {Centre de Nanosciences et de Nanotechnologies, CNRS, Univ. Paris-Sud, Universit\'{e} Paris-Saclay, C2N---Marcoussis, 91460 Marcoussis, France}

\author      {A. Amo}
\affiliation {Centre de Nanosciences et de Nanotechnologies, CNRS, Univ. Paris-Sud, Universit\'{e} Paris-Saclay, C2N---Marcoussis, 91460 Marcoussis, France}

\author{C. Ciuti}
\affiliation {Laboratoire Mat\'eriaux et Ph\'enom\`enes Quantiques,
Universit\'{e} Paris Diderot, Sorbonne Paris Cit\'{e} and CNRS, UMR
7162,  75205 Paris Cedex 13, France}

\author      {J. Bloch}
\affiliation {Centre de Nanosciences et de Nanotechnologies, CNRS, Univ. Paris-Sud, Universit\'{e} Paris-Saclay, C2N---Marcoussis, 91460 Marcoussis, France}

\date{\today}

\begin{abstract}
We experimentally explore the dynamic optical hysteresis of a semiconductor microcavity as a function of the sweep time.  The hysteresis area exhibits a double power law decay due to the shot noise of the driving laser, which triggers switching between metastable states. Upon increasing the average photon number and approaching the thermodynamic limit, the double power law evolves into a single power law. This algebraic behavior characterizes a dissipative phase transition. Our findings are in good agreement with theoretical predictions, and the present experimental approach is promising for the exploration of critical phenomena  in photonic lattices.
\end{abstract}

\narrowtext

\maketitle

Optical bistability --- the existence of two stable states with different photon numbers for the same driving conditions ---  is a general feature of driven nonlinear systems described within the mean-field approximation (MFA)~\cite{Gibbs}. Beyond the MFA, a quantum treatment predicts that the steady-state of a nonlinear cavity is unique at any driving condition~\cite{WallsMilburn}. The origin of this apparent contradiction was  noted by Bonifacio and Lugiato~\cite{Bonifacio78}, and by Drummond and Walls~\cite{Drummond80}: quantum fluctuations (the lost feature in the MFA) trigger switching between states and the exact solution corresponds to a weighted average over the two metastable states. Experiments in the 80's with two-mode lasers  evidenced extremely long switching times~\cite{Roy80}, which were predicted to diverge for weak fluctuations and/or large photon numbers~\cite{Hioe81}. Already in these early works, this dramatic slowing down of the system dynamics was linked to a first order phase transition~\cite{Roy80, Lett81, Hioe81}.

The physics of nonlinear resonators is receiving renewed interest in connection to predictions of quantum many-body phases~\cite{Hartmann06, Greentree06, Angelakis,  LeBoite, Wilson16, Biondi16}, critical phenomena~\cite{ Carmichael15, Casteels16,  Mendoza16, Wilson16, Casteels16arxiv, Foss16, Biondi16}, and  dissipative phase transitions~\cite{Kessler12}. Impressive progress is being made in building lattices of nonlinear resonators, such as photonic crystal cavities~\cite{Vuck12, Hamel15}, waveguides~\cite{Fleischer03}, superconducting microwave resonators\cite{Vijay09, Underwood12}, or optomechanical resonators~\cite{Painter09, Favero16}. In this context, semiconductor microcavities operating in the exciton-photon strong coupling regime provide a versatile platform where photon hopping and the pumping geometry can be controlled~\cite{CarusottoRMP}. Lattices of different dimensionalities can be engineered~\cite{Kim11, Baboux16}, and the hybrid light-matter nature of their elementary excitations, namely cavity polaritons, provide a strong and tunable Kerr nonlinearity via the exciton component~\cite{Amo09, Baas04, Paraiso10, Rodriguez16}.

Recently, it was predicted that even in a single resonator, critical exponents could be retrieved from dynamical hysteresis measurements~\cite{Casteels16arxiv}. More precisely, when the driving power is swept at a finite speed across a bistability, the area of the hysteresis cycle is expected to close following a double power-law as a function of the sweep time~\cite{Luse94, Casteels16}. The long-time decay arises from quantum fluctuations, and presents a universal $-1$ exponent~\cite{Casteels16}. In the thermodynamic limit wherein the photon number in the bistability tends to infinity and fluctuations are negligible, the algebraic decay of the hysteresis area is expected to evolve into a single power law~\cite{Casteels16arxiv}. This behavior characterizes a first order dissipative phase transition~\cite{Casteels16arxiv}. Exploring such dissipative phase transitions in a single cavity  is a key step towards future studies with more complex systems.

In this Letter, we experimentally demonstrate the algebraic decay of the dynamical optical hysteresis in semiconductor micropillars. Scanning the power up and down at decreasing speeds, we observe the progressive closure of the hysteresis cycle induced by quantum fluctuations.  The hysteresis area shows a double power-law decay as a function of the sweep time, with experimentally retrieved exponents in agreement with theoretical predictions. Probing different laser detunings and photon-photon interactions, we show that the algebraic decay evolves towards a single power law when the photon number becomes very large, i.e. when approaching the thermodynamic limit. Our results pave the way to the investigation of dissipative phase transitions in lattices of nonlinear resonators.

\begin{figure}[!]
 \centerline{{\includegraphics[width=\linewidth]{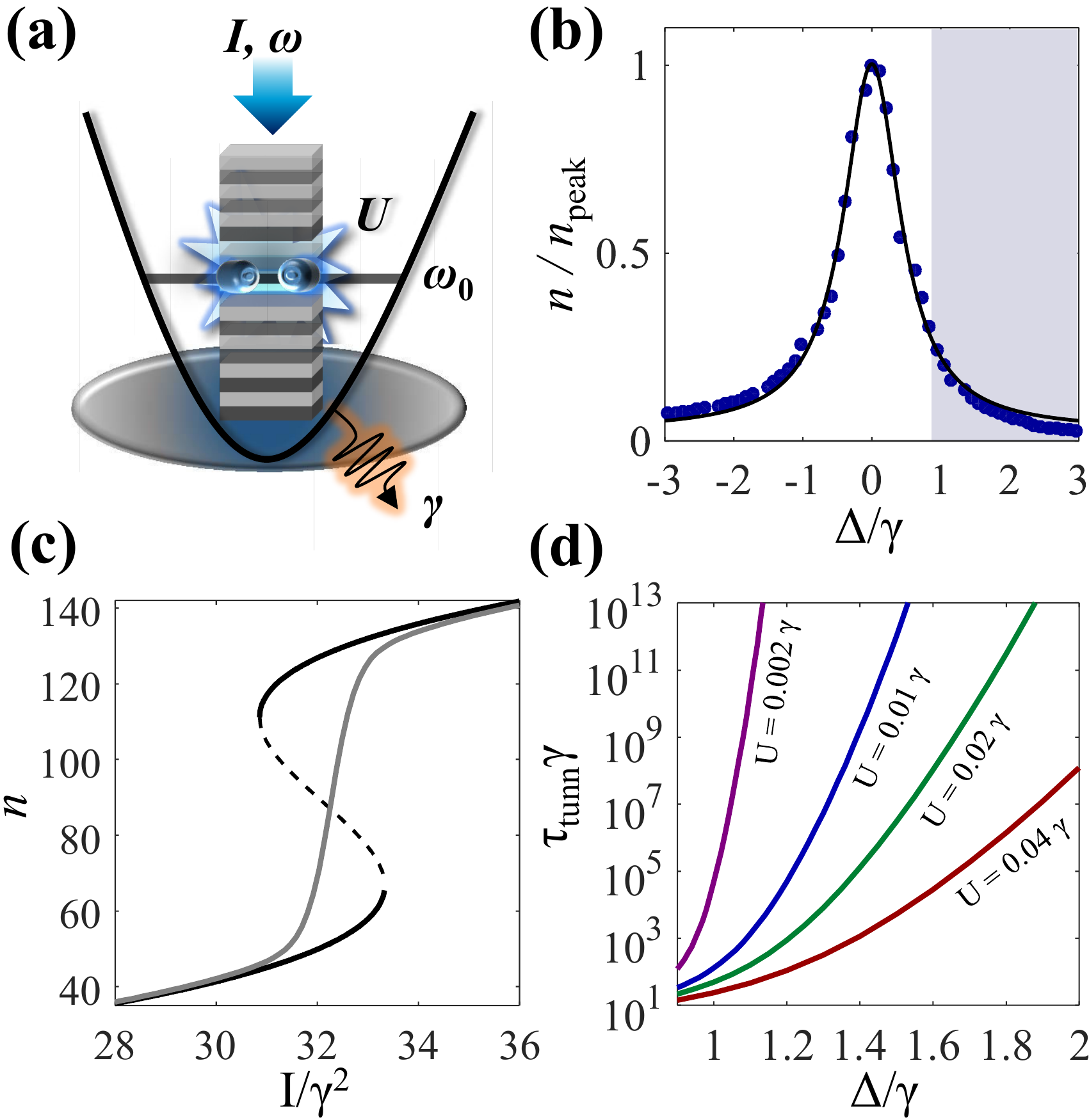}}}\caption{(a) Sketch of a microcavity with mode frequency $\omega_0$, loss rate $\gamma$, photon-photon interactions of strength $U$, driven by an electromagnetic field of intensity $I$ and frequency $\omega$. (b) Normalized photon density in a semiconductor microcavity under weak driving. Experimental data points for the sample studied in Figs.~\ref{fig2},~\ref{fig3}, and~\ref{fig4}  are fitted with a Lorentzian lineshape. The shaded area indicates the mean-field bistable regime $\Delta \equiv\omega-\omega_0 > \sqrt{3} \gamma/2$ for $U>0$.  (c) Mean-field (black curves) and quantum (gray curves) solutions for a cavity with $U = 0.0075$ $\gamma$ probed at $\Delta=\gamma$.  In the mean-field solution the solid and dashed curves are stable and unstable states, respectively.(d) Tunneling time $\tau_{\textrm{tunn}}$ between the two mean-field stable solutions.}
\label{fig1}
\end{figure}

First, we  briefly revisit the  physics of  a driven-dissipative single mode nonlinear cavity as illustrated in Fig.~\ref{fig1}(a). $\omega_0$, $\gamma$,  and  $U$ represent the mode frequency, loss rate, and photon-photon interaction strength (Kerr nonlinearity) of the cavity, driven by an electromagnetic field of frequency $\omega$ and intensity $I$. Within the rotating-wave approximation, the Hamiltonian (in units of $\hbar = 1$) is:

\begin{equation}
\hat{H}(t) = \omega_0 \hat{a}^\dagger\hat{a}+\frac{U}{2}\hat{a}^\dagger\hat{a}^\dagger\hat{a}\hat{a} + \sqrt{I}\left(e^{-i\omega t}\hat{a}^{\dagger} + e^{i\omega t}\hat{a}\right).
\label{Eq: SysHam}
\end{equation}

\noindent The boson operator $\hat{a}$ ($\hat{a}^{\dagger}$) annihilates (creates) an excitation in the resonator. The dynamics is described by the Lindblad master equation  for the density matrix $\hat{\rho}(t)$:

\begin{eqnarray}
\frac{\partial\hat{\rho}(t)}{\partial t}= & i\left[\hat{\rho},\hat{H}(t)\right] + \frac{\gamma}{2}\left( 2\hat{a}\hat{\rho}\hat{a}^\dagger - \hat{a}^\dagger\hat{a}\hat{\rho} - \hat{\rho}\hat{a}^\dagger\hat{a} \right).
\label{eq:Master}
\end{eqnarray}

Equation~\ref{eq:Master} can  be written  as $\partial_t\hat{\rho}
=\hat{\mathcal L}\hat{\rho}$, where $\hat{\mathcal L}$ is the Liouvillian
superoperator. $\hat{\mathcal L}$ has a
complex spectrum, of which two eigenvalues $\lambda$ are particularly relevant for the long-time dynamics: i)  $\lambda = 0$ corresponds to the steady-state, and ii) the  non-zero eigenvalue with real part closest to zero is  the Liouvillian gap $\bar{\lambda}$.

An exact expression for the steady-state photon density predicted by Eq.~\ref{eq:Master} was found in Ref.~\onlinecite{Drummond80}. This exact solution is shown as a gray line in Figure~\ref{fig1}(c), for $U/\gamma = 0.0075$ and a laser-cavity detuning $\Delta = \omega - \omega_0 = \gamma$.  The MFA follows from assuming the field to be coherent with amplitude $\alpha (t) = \langle \hat{a} \rangle$. Equation \ref{eq:Master}  then reduces to:   $i\frac{\partial \alpha}{\partial t} = \left(\omega_0 - i \frac{\gamma}{2} + U \left|\alpha \right|^2\right) \alpha + \sqrt{I} e^{-i\omega t}.$ The black line in Fig.~\ref{fig1}(c) is the corresponding MFA calculation, displaying bistability for $31<I/\gamma^2<33$. While the  MFA implies a hysteresis cycle when varying the power across the bistability, the quantum solution is unique.  This apparent contradiction is due to the absence of fluctuations in the MFA~\cite{Bonifacio78, Drummond80}. Fluctuations  (quantum or classical) render the mean-field steady-states  metastable~\cite{Mabuchi11, Deveaud15}, and  the exact quantum solution corresponds to their average.

\begin{figure*}[!] \centerline{{\includegraphics[width=\linewidth]{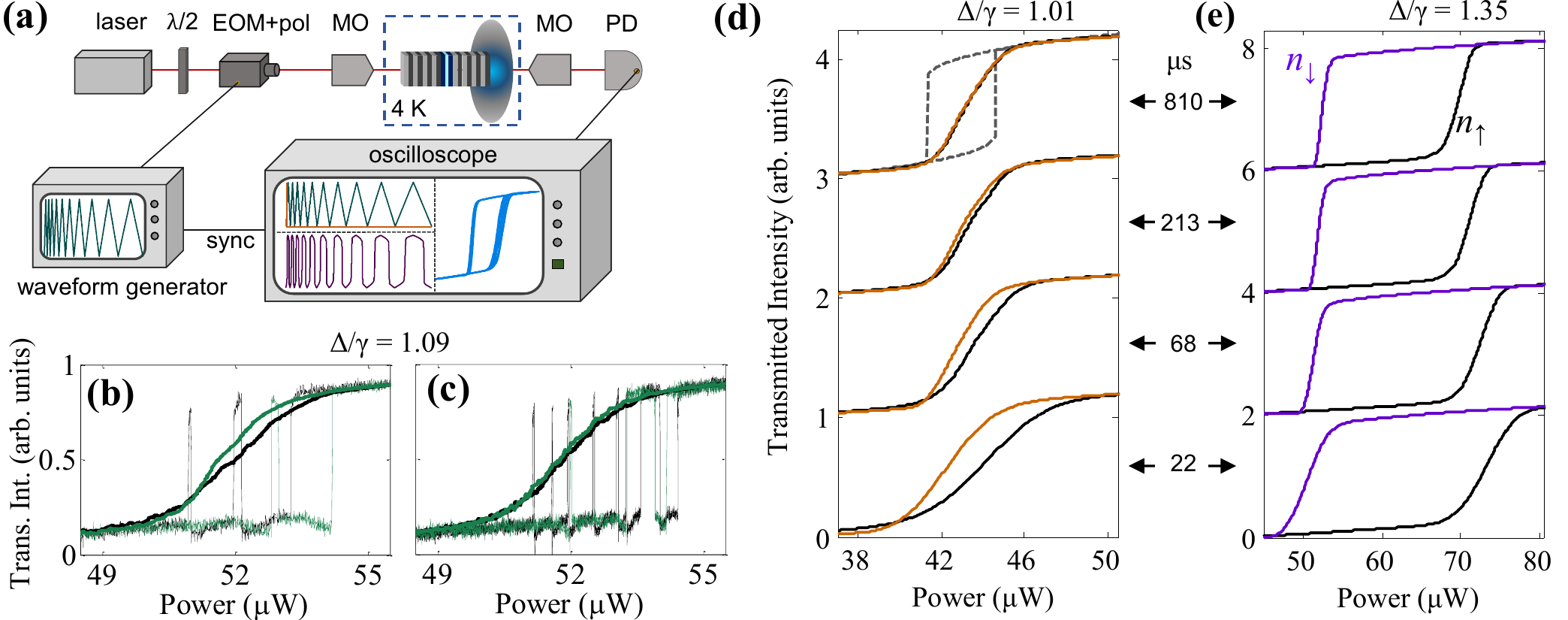}}}\caption{(a) Experimental setup: $\lambda/2$, MO, PD, and  EOM+pol, stand for half-wave plate, microscope objective, photodiode, and electro-optic modulator with a polarizer, respectively.  The green (purple) traces in the waveform generator and in the oscilloscope are measurements of the incident (transmitted) signals. The hysteresis cycles in the oscilloscope are obtained by plotting the transmitted versus the incident signal, overlaid for various scanning times.  The colored and black lines in (b)-(e) represent the transmission when the  power is ramped down and up, respectively. (b) and (c) show single shot (thin lines) and averages over 1000 realizations (thick lines) of dynamic hysteresis. The scanning time is $t_s= 0.11$ ms in (b), and  $t_s= 0.43$ ms in (c). (d) and (e) show dynamic hysteresis averaged over 1500 realizations.  The dashed line  in (d) is  the mean-field calculation corresponding to the experiment. } \label{fig2} \end{figure*}

The reconciliation between numerous reports of optical bistability~\cite{Gibbs76, Dorsel83, Mandel90, Rempe91, Collot93, Baas04,  Lipson04, Notomi05, Wurtz06, Boulier14}  and the quantum prediction of a unique steady-state~\cite{Drummond80} follows from the fact that fluctuations can take astronomical times to induce switching between  metastable states. Historically, this switching time is known as the tunneling time for bistability $\tau_{\textrm{tunn}}$~\cite{Risken87, Vogel88},  first-passage time~\cite{Roy80}, quantum activation time~\cite{Dykman12}, or (inverse) asymptotic decay rate~\cite{Kessler12}.  We will label this characteristic time as $\tau_{\textrm{tunn}}$,  which is the longest reaction time of the system given by the minimum value of $\bar{\lambda}$. Figure~\ref{fig1}(d) shows $\tau_{\textrm{tunn}}$ as a function of $\Delta/\gamma$ for different $U/\gamma$.  For weak interactions and/or large detunings,  $\tau_{\textrm{tunn}}$ can vastly exceed realistic measurement times.  Consequently, hysteresis measurements performed within a shorter time than $\tau_{\textrm{tunn}}$ lead to an apparent bistability. In this vein, Casteels and co-workers predicted how the hysteresis area should be influenced by  quantum fluctuations when  the scanning time across the ``bistability'' is commensurate with  $\tau_{\textrm{tunn}}$~\cite{Casteels16}. In particular, they predicted double power-law decay of the hysteresis area~\cite{Casteels16}, in contrast with previous reports of a single power-law decay~\cite{Mandel90}.

To measure dynamic optical hysteresis, we use rectangular micropillars etched from a GaAs $\lambda$ planar cavity containing one 8 nm In$_{0.04}$Ga$_{0.96}$As quantum well and surrounded by two Ga$_{0.9}$Al$_{0.1}$As/Ga$_{0.05}$Al$_{0.95}$As distributed Bragg reflectors with 26 and 30 pairs of layers at the top and bottom, respectively. The sample is maintained at 4 K and driven by a frequency-tunable single-mode laser. We probe the lowest energy mode of the micropillars, whose linewidth ranges from 28 to 34 $\mu$eV~\cite{supp}. The value of $U$  is estimated from the energy of the confined polariton mode and its exciton fraction~\cite{supp}. The laser  power is modulated by an electro-optic modulator (EOM) fed by a waveform generator [see Figure~\ref{fig2}(a)]. The waveform contains a series of $\sim 50$ triangular ramps of variable time-duration. The transmission  through the cavity is measured with a photodiode connected to an oscilloscope.  The scanning times $t_s$ (the time it takes to ramp the power from the lowest to the highest value) span the 0.8-50 kHz range. As shown in the supplemental information, laser shot noise is the only noise source within this frequency range and we exclude additional fluctuations from our observations~\cite{supp}.

We are interested in the hysteresis area,

\begin{equation}
A = \int_{P_{\mathrm{min}}}^{P_{\mathrm{max}}} |n_{\downarrow}(P) - n_{\uparrow}(P)|  \mathrm{d}P,
\label{eq:A}
\end{equation}

\noindent  as a function of $t_s$.  $n_{\downarrow}(P)$ and $n_{\uparrow}(P)$ represent the cavity transmission when the power is ramped down and up, respectively. $P_{\mathrm{min}}$ and $P_{\mathrm{max}}$ are powers below and above the hysteresis range.   In the absence of fluctuations, $A$ saturates to a finite value (the mean-field static hysteresis area) for  $t_s \to \infty$~\cite{Mandel90}. However, unavoidable quantum fluctuations induce spontaneous switchings between the mean-field ``bistable'' states. Consequently, the hysteresis area averaged over many realizations, $A_{\textrm{av}}$, is expected to close in proportion to the number of switching events,  such that $\lim_{t_s \to \infty} A_{\textrm{av}}=0$.

In Figs.~\ref{fig2}(b, c) we compare single-shot (thin lines) and averaged (thick lines, 1000 realizations) transmission measurements for $\Delta/\gamma=1.09$.  The single-shot measurements in Fig.~\ref{fig2}(c) display  more switchings than in Fig.~\ref{fig2}(b) because the sweep is slower in Fig.~\ref{fig2}(c).  Consequently, the average hysteresis area in Fig.~\ref{fig2}(c) is reduced.

Figure~\ref{fig2}(d) show hysteresis measurements (averaged over 1500 realizations) for $\Delta/\gamma=1.01$ and different values of $t_s$. The hysteresis area close for increasing $t_s$. For the slowest sweep, the measured cycle strongly deviates from the mean-field prediction (dashed lines) and resembles the exact quantum prediction in Fig.~\ref{fig1}(c). In contrast, for larger $\Delta/\gamma$, $\tau_{\textrm{tunn}} \gg t_s$ and the hysteresis area changes marginally [see Fig.\ref{fig2}(e)].

\begin{figure} \centerline{{\includegraphics[width=\linewidth]{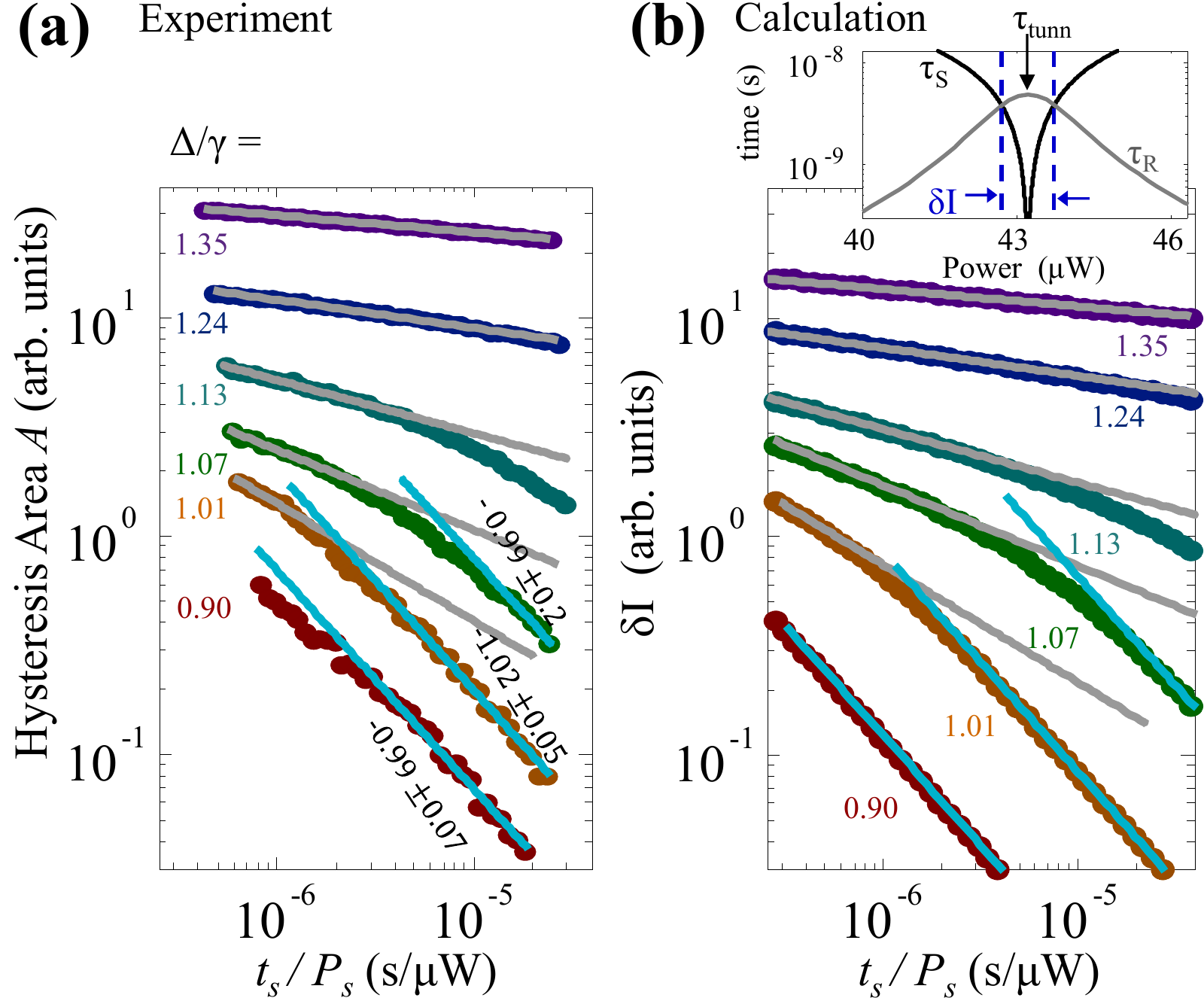}}}\caption{(a) Measured hysteresis area $A_{\textrm{av}}$ (defined in Eq.~\ref{eq:A}) as a function of $t_s/P_s$, where  $P_s$ is the scanned power range. Different colors correspond to different values of $\Delta/\gamma$. The gray lines are power law fits with an exponent greater than -1. The blue lines indicate power laws in the regime influenced by quantum fluctuations. The experimentally retrieved exponents in this regime are shown with $2\sigma$ confidence intervals on the fits. (b) Calculations of the non-adiabatic range $\delta I$ of the driving intensity  using the scaling analysis described in the text. The power laws in (b) all have the same exponents retrieved from the fits in (a). The inset in (b) shows the system reaction time $\tau_R$ in gray, and the sweep time scale $\tau_S$ in black, for $U = 0.0075$ $\gamma$ and $\Delta = 1.01$ $\gamma$. The dashed blue lines indicate the  non-adiabatic range $\delta I$.  The conversion of the theoretical intensity units to the experimental power units is described in the supplemental information~\cite{supp}.}
\label{fig3}
\end{figure}

The behavior of $A_{\textrm{av}}$ not only depends on $t_s$, but also on the scanned power range $P_s \equiv P_{\mathrm{max}} - P_{\mathrm{min}}$ .  The ratio $t_s/P_s$ gives an effective (inverse) sweep speed.  In Fig.~\ref{fig3}(a) we plot $A_{\textrm{av}}$ as a function of $t_s/P_s$ for 6 different $\Delta/\gamma$. For small $\Delta/\gamma$ we observe two power laws indicated by the gray and blue lines in Fig.~\ref{fig3}(a). The blue lines correspond to a power-law with a $-1$ exponent, as predicted in Ref.~\onlinecite{Casteels16}. This is the regime dominated by quantum fluctuations, when $\tau_{\textrm{tunn}} < t_s$.  For increasing $\Delta/\gamma$, the average photon number in the bistability increases and fluctuations become relatively weaker. Consequently, the $-1$ power law sets in at longer times and $A_{\textrm{av}}$ follows a single power law within the experimental observation window.

Calculating the dynamics behind the results in Fig.~\ref{fig3}(a) requires a time-evolution  up to $10^8$ times the polariton lifetime ($21$ ps), a temporal resolution below the polariton lifetime, and a dimensionality of the Hilbert space of $\sim10^3$. To circumvent this difficulty, Ref.~\onlinecite{Casteels16} introduced a method  based on a scaling analysis in the spirit of the Kibble-Zurek mechanism for dynamic  phase transitions~\cite{Jacek10}. The key idea is that a power sweep at a finite rate across the bistability  involves  a non-adiabatic response of the system, resulting in the dynamic hysteresis. The non-adiabatic intensity range $\delta I$  is determined by comparing the  sweep time scale $\tau_S$ with the system reaction time $\tau_R$  (see Fig. \ref{fig3} (b) inset and supplementary information~\cite{supp}). The longest reaction time is $\tau_{\textrm{tunn}}$. Similar to the hysteresis area $A_{\textrm{av}}$, $\delta I$ exhibits a double power-law as a function of the sweep rate~\cite{Casteels16}.

Figure~\ref{fig3}(b) shows calculations of $\delta I$ reproducing our experimental observations. The lines in Fig.~\ref{fig3}(b) all have slopes deduced from the power law fits to the measurements in Fig.~\ref{fig3}(a). A good agreement between measurements and calculations is obtained for every $\Delta/\gamma$. This confirms the exponents expected for a single mode cavity under the influence of quantum fluctuations. To obtain this good agreement, we adjusted the value of $U/\gamma$ within the experimental uncertainty. We take $U/\gamma=7.5 \cdot 10^{-3}$, whereas the experimental estimate~\cite{supp} is  $U/\gamma=2 \cdot 10^{-3} \substack{+8 \cdot 10^{-3} \\ -1.6 \cdot 10^{-3}}$. Overall, the results in Fig.~\ref{fig3} show that as $\Delta/\gamma$ decreases and the photon number in the bistability decreases, the hysteresis area evolves from a single to a double power law decay. This transition is due to the influence of quantum fluctuations.

\begin{figure}\centerline{{\includegraphics[width=\linewidth]{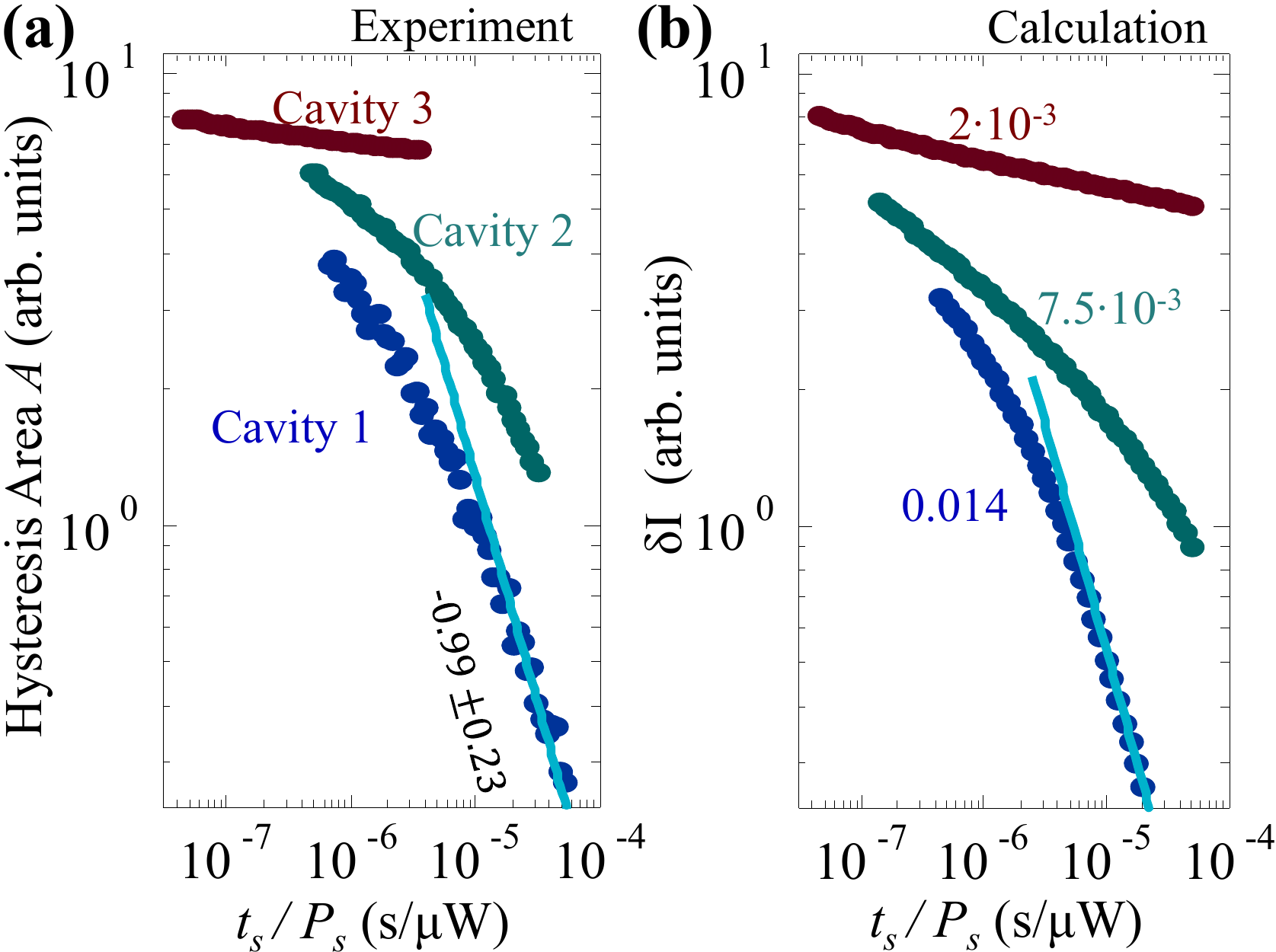}}}\caption{(a) Measurements of the hysteresis area $A$ for three cavities with different $U/\gamma$ and approximately equal $\Delta/\gamma$.  $U/\gamma$ decreases from cavity 1 to cavity 3. (b) Calculations of $\delta I$ as explained in the text and in Fig.~\ref{fig3}. For the highest and lowest curves  $\Delta/\gamma =1.15\pm0.1$, while for middle curve $\Delta/\gamma=1.13 \pm0.1$, both in experiments and calculations.  In (a), the curve corresponding to cavity 3 was divided by 80. In (b), the curve corresponding to the smallest $U/\gamma$ was divided by 4, and the curve  corresponding to the largest $U/\gamma$ was multiplied by 2. These multiplications (for improving visibility) only shift the curves vertically and do not change the exponent.} \label{fig4}
\end{figure}

A thermodynamic limit can be defined for a single resonator by letting $N \to \infty$ and $U \to 0$ while keeping $U\cdot N$ constant~\cite{Casteels16arxiv}. This limit can be explored probing cavities with different values of $U/\gamma$ at a fixed laser-cavity detuning  $\Delta/\gamma$. Experimentally, we vary $U/\gamma$ by selecting micropillars with different lateral dimensions. A reduced cross-sectional area of the micropillar blue-shifts the energy of the confined polariton modes and increases their exciton fraction, thereby increasing $U/\gamma$~\cite{supp}.   Figure~\ref{fig4}(a) shows measurements of $A_{\textrm{av}}$  for three cavities probed at  $\Delta/\gamma = 1.15\pm 0.1$. For cavity 1 with the strongest interaction strength,  $A_{\textrm{av}}$ displays a double power law with the -1  exponent at large $t_s/P_s$.  As $U/\gamma$ decreases, the time at which the power law with the $-1$ exponent sets in increases. For cavity 3 with the weakest interaction strength, $A_{\textrm{av}}$ depends marginally on $t_s/P_s$ and the data follows a single power law. These observations are consistent with the dramatic dependence of $\tau_{\textrm{tunn}}$ on $U/\gamma$  plotted in Fig.~\ref{fig1}(d).

Figure~\ref{fig4}(b) shows calculations based on the scaling analysis previously described, in good agreement with the measurements in Fig.~\ref{fig4}(a). Details about the values of  $U/\gamma$ used in the calculations are discussed in the supplemental information~\cite{supp}.  Overall,  Fig.~\ref{fig4} demonstrates that as $U\to0$ and the average photon number in the bistability increases,  the hysteresis area evolves towards a single power-law decay. This is the signature of a system approaching the thermodynamic limit of high photon numbers~\cite{Casteels16arxiv}.

To summarize, we showed a double power-law decay of the hysteresis area as a function of the sweep rate. As the average photon number increases and quantum fluctuations become relatively weaker, the tunneling time increases dramatically. This shifts the transition to the power law with exponent -1 to larger times. In the thermodynamic limit of large photon numbers, the hysteresis area exhibits a single power-law decay. These results open the way to the exploration of dissipative phase transitions in lattices of coupled micropillars, where photon hopping can give rise to intriguing behavior. For instance, a square lattice of bistable resonators has been  mapped to an equilibrium Ising model with an effective temperature given by the losses~\cite{Foss16}. The question remains open regarding phase transitions in more elaborate lattices with intricate topologies~\cite{Jacqmin}, with spin-orbit coupling~\cite{Sala}, or with quasi-crystalline structure~\cite{Tanese14} in which thermodynamic properties  reflect their non-integer dimensions~\cite{Akkermans09, Akkermans10}.

\begin{acknowledgments}
This work was supported by the Marie Curie individual fellowship PINQUAR (Project No. 657042), the French National Research Agency (ANR) program Labex NanoSaclay via the projects Qeage (ANR- 11-IDEX-0003-02) and ICQOQS (ANR-10-LABX-0035), the French RENATECH network, the ERC grant Honeypol and the EU-FET Proactiv grant AQUS (Project No. 640800). W.C., F. S., and C.C. acknowledge support from ERC (via the Consolidator
Grant "CORPHO" No. 616233.
\end{acknowledgments}

\clearpage
\section{Supplemental Material}

\noindent\textbf{Sample details}

The planar cavity was grown by molecular beam epitaxy and comprises a  GaAs $\lambda$ cavity between two Ga$_{0.9}$Al$_{0.1}$As/Ga$_{0.05}$Al$_{0.95}$As distributed Bragg reflectors with 26 and 30 pairs of layers at the top and bottom, respectively.   One 80 \r{A}-wide In$_{0.04}$Ga$_{0.96}$As quantum well is positioned at the center of the cavity.  A Rabi splitting of  $3.4$ meV results from strong exciton-photon coupling. The three rectangular micropillars (all having 2:1 aspect ratio) were fabricated by electron beam lithography and dry etching of the planar cavity. With reference to the labels in Fig. 4(a) of the main manuscript, the cross-sectional area  $A_c$ of the micropillars are:  $A_c = 6$ $\mu \mathrm{m}^2$ for cavity 1,  $A_c = 8$  $\mu \mathrm{m}^2$  for cavity 2, and $A_c = 29$ $\mu \mathrm{m}^2$ for cavity 3. \\

\noindent\textbf{Retrieval of parameters}

The mode frequency $\omega_0$ and polariton loss rate $\gamma$ of each micropillar were deduced by measuring the transmitted spectrum under weak driving ($<1 \mu$W), and fitting a Lorentzian lineshape as shown in Fig. 1(b) of the main manuscript.  The polariton-polariton interaction energy $U$ of each micropillar was estimated following the procedure described in detail in Ref.~\onlinecite{Rodriguez16}, and which we summarize below. To begin, we took the exciton-exciton interaction constant to be $g_{\mathrm{exc}} = 30$ $\mu$eV $\cdot$  $\mu \mathrm{m}^2$. This value is consistent with theoretical predictions~\cite{Ciuti98} and with our previous observations~\cite{Rodriguez16}. Next, we calculated the 2D  polariton-polariton interaction constant as $U_{2\mathrm{D}} = g_{\mathrm{exc}} |X|^4$, with   $|X|^2$ the exciton fraction of the polariton admixture.  To obtain the value of $|X|^2$ as a function of energy, we analyzed the exciton-polariton dispersion in an effectively 2D cavity adjacent to the micropillars in the wafer. In particular,  we fitted the eigenvalues of a $2 \times 2$  Hamiltonian to the polariton dispersion, and obtained $|X|^2$ from the eigenvector associated with the lower polariton eigenvalue (energy).  We then evaluated the value of $|X|^2$ at the energy of each micropillar mode to calculate $U_{2\mathrm{D}}$.  Finally, using the cross-sectional area of each micropillar $A_c$, we obtain $U= U_{2\mathrm{D}}/A_c$.

The retrieved parameters are:

\noindent i) For the cavity under study in all figures of the main manuscript [cavity 2 in Fig. 4(a)], we have $\hbar \omega_0 = 1478.687 \pm 0.004$ meV, $\gamma=31 \pm 2 \mu eV$, $|X|^2=0.13$, and  $U/\gamma=2  \substack{+8 \ \\ -1.6 } \cdot 10^{-3}$.

\noindent  ii) For cavity 1 in Fig. 4(a) of the main manuscript, we have $\hbar \omega_0 = 1479.695 \pm 0.004$ meV, $\gamma=31 \pm 3 \mu eV$, $|X|^2=0.21$,  and $U/\gamma=1 \substack{+4  \\ -0.8 } \cdot 10^{-2}$ \\

\noindent   iii) For cavity 3 in Fig. 4(a) of the main manuscript, we have  $\hbar \omega_0 = 1472.752 \pm 0.004 $ meV,$\gamma=30 \pm 2 \mu eV$,  $|X|^2=0.02$, and $U/\gamma=2  \substack{+8 \ \\ -1.6 } \cdot 10^{-5}$.\\

\noindent\textbf{Measurement details}

The  driving laser is a tunable MSquare Ti:Sapphire oscillator with a linewidth below $10$ MHz. The laser frequency is locked with an accuracy of 0.1 pm. Given the typical values of $\gamma$ and $\hbar \omega_0$ (see above), this translates to an uncertainty in $\Delta/\gamma$ on the order of  $\pm 0.01$.  The excitation and collection objectives have a numerical aperture of 0.5 and 0.4, respectively. For all three cavities the excitation laser beam is linearly polarized parallel to the long axis of the rectangular micropillars, thereby probing the lowest energy mode.  For the cavity under study in all figures of the main text, the lowest energy mode for the orthogonal polarization  (along the short axis) is $\sim 5 \gamma$ higher in energy, while co-polarized higher energy states are several tens of $\gamma$ away.   Thus, the single mode approximation holds reasonably well.

To verify that our dynamic hysteresis measurements were shot-noise limited, we analyzed the power spectral density of the driving laser. The laser beam passed through our entire optical setup to reproduce the experimental conditions of the dynamic hysteresis measurements. At low frequencies (1- 50 Hz), where various fluctuations are significant, Fig.~\ref{fig1}(a) shows that the noise power scales linearly with the total power. In contrast, at higher frequencies Fig.~\ref{fig1}(b) shows that the noise power scales with the square root of the total power, as expected for shot noise. We also observed a frequency-independent noise power spectral density (characteristic of shot noise) in the frequency range of Fig.~\ref{fig1}(b). In addition, we note that the power in the modulated signal of our dynamic hysteresis measurements exceeds the noise level by  more than four orders of magnitude. \\

\begin{figure}[!] \centerline{{\includegraphics[width=\linewidth]{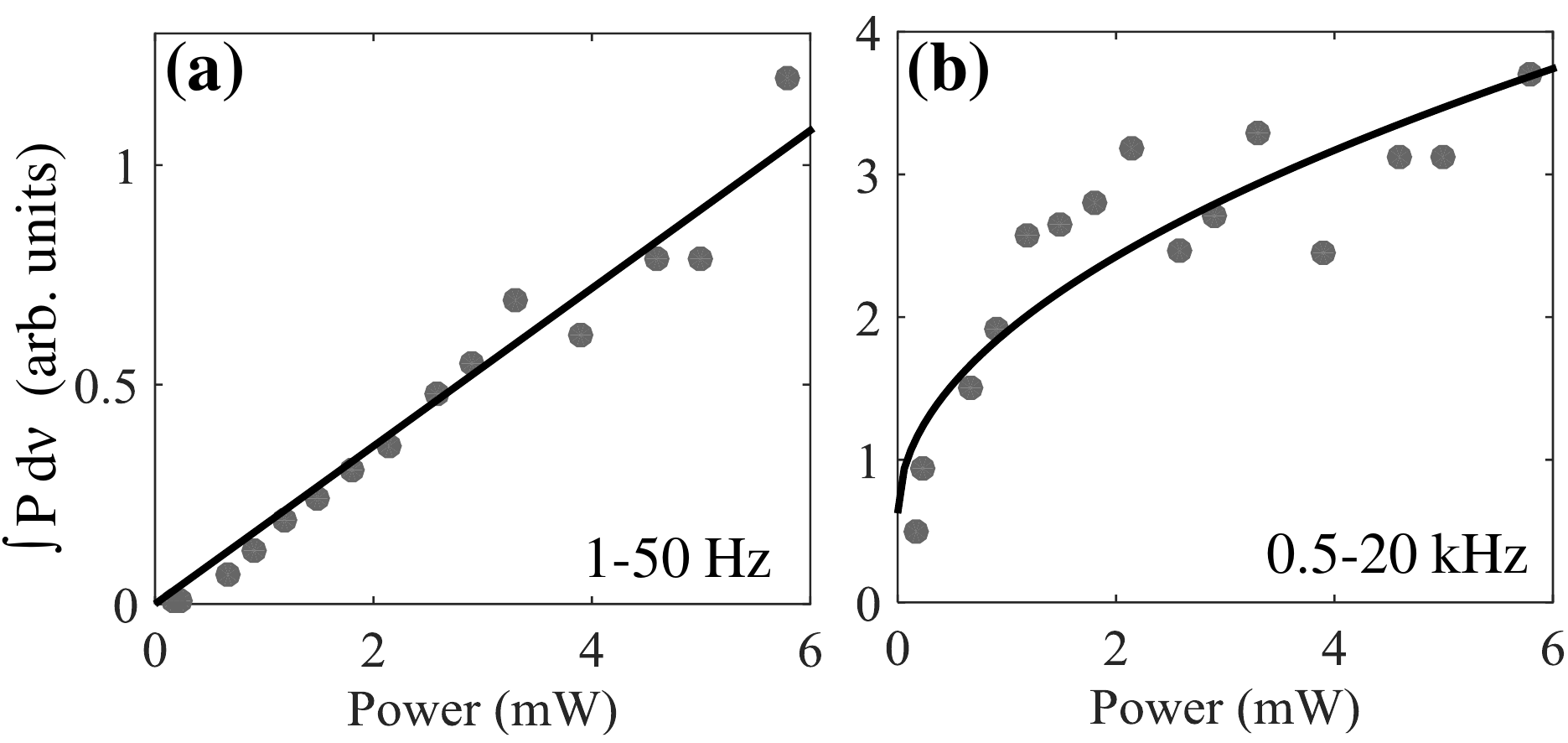}}}\caption{Power-dependence of the noise in the driving laser  integrated over the frequency range indicated in each panel. The black line in (a) is a linear fit, and the black line in (b) is a square root fit.  The square root behavior at high frequencies is indicative of shot noise. }
\label{fig1}
\end{figure}

\noindent\textbf{Noise sources}

Additional sources of noise (besides shot noise) could also induce switching between metastable states. For instance, in Ref.~\onlinecite{Deveaud15} a large amount of noise was deliberately added to the driving laser to trigger switching between the two metastable states. In addition, noise could be generated within the cavity. For example, the intracavity polariton field could be subjected to fluctuations linked to acoustic phonon scattering or two photon absorption.  In our experiments,  phonon scattering can be excluded because $kT$ is much smaller that the energy difference between the state we probe and the next confined polariton state. Fluctuations induced by two photon absorption can  be ruled out considering the experiments reported in Fig.4(a). When  probing smaller values of $U / \gamma$ for fixed $\Delta / \gamma$, the driving intensity at the critical point is larger. Stronger driving is expected to make two-photon absorption more relevant and thereby reduce the hysteresis area. However, the opposite behavior is observed and we can therefore neglect additional fluctuations induced within the cavity.   Thus, we can safely consider that shot noise is the only relevant source of noise in our experiments.\\

\noindent\textbf{Mean-field calculation details}

For the dynamic hysteresis calculations, we consider a triangular modulation of the drive intensity consisting of a linear sweep from $I_0$ to $I_0 + P_s$ followed by the reverse one from $I_0 + P_s$ back to $I_0$:
\begin{equation}
I(t) = I_0 + \frac{t}{t_s}P_s \theta(t_s-t) - \frac{t - 2t_s}{t_s}P_s \theta(t - t_s),
\label{eq:Sweep}
\end{equation}
Here, $P_s$ is the range of the power sweep, $t_s$ is the scanning time, and the effective sweep velocity is $P_s/t_s$. Introducing the sweep (\ref{eq:Sweep}) in the  equation of motion for the mean-field $\alpha$ (see main manuscript) with $t_s/P_s = 10^6 /(5\gamma^3) = 2.1\cdot 10^{-5} s/\mu$W leads to the dynamic hysteresis presented in Fig. 2(d) of the main manuscript.\\

\noindent\textbf{Scaling Analysis}

In general, a master equation can be defined in terms of its Liouvillian
superoperator $\hat{\mathcal L}$, namely $\partial_t\hat{\rho}
=\hat{\mathcal L}\hat{\rho}$. The Liouvillian superoperator has a
complex spectrum of eigenvalues $\lambda$.  Their imaginary part has the meaning of an
excitation frequency, while their real part determines the dissipation
rate. The steady-state density
matrix $\hat{\rho}_{ss}$ corresponds to the eigenvector of
$\hat{\mathcal L}$ with eigenvalue $\lambda = 0$, namely $\hat{\mathcal
L} \hat{\rho}_{ss} = 0$. The dynamical properties depend on the non-zero
complex eigenvalues. The Liouvillian gap $\bar{\lambda}$ is defined as
the non-zero eigenvalue of $\hat{\mathcal L}$ with real part closest to
zero. In general, the Liouvillian gap becomes highly suppressed as a
system enters a critical region~\cite{Kessler12}. This leads to a
critical slowing down of the dynamics, and a dissipative phase
transition when the gap $\bar{\lambda}$ closes. Around the bistability
the imaginary part of $\bar{\lambda}$ is strictly zero and $\vert
\text{Re} \bar{\lambda} \vert$ exhibits a minimum at a driving intensity
in the midst of the bistability region~\cite{Casteels16}.  From the real
part of $\bar{\lambda}$ the reaction time $\tau_R$ can be determined as:
$\tau_R = 1/\vert \text{Re}\bar{\lambda} \vert$. $\tau_R$ corresponds to
the longest timescale on which the system relaxes to its steady-state.
The transition point is defined as the intensity $I_t$ where $\tau_R$
reaches the maximal value $\tau_{\textrm{tunn}}$, the tunneling time.

A consequence of the above behavior of the Liouvillian gap is that the system can respond non-adiabatically when the driving power is varied across the optical bistability region. The sweep timescale $\tau_S$ is defined by  the inverse normalised transition rate of the sweep (\ref{eq:Sweep}): $\tau_S = |\dot{\epsilon}/\epsilon|^{-1}$, where $\epsilon$ is the distance from the transition point: $\epsilon = I(t) - I_t$. The non-adiabatic range is reached when the sweep timescale  $\tau_S$ is smaller than the system reaction time $\tau_R$, i.e. for $\tau_S < \tau_R$. If this condition is fulfilled, the system can not relax to its steady-state and a dynamic hysteresis arises. The point where the system enters or exits this non-adiabatic regime can be estimated by equating the two timescales: $\tau_S = \tau_R$. This allows to determine the size of the non-adiabatic range $\delta I$, as indicated by the blue dashed lines in the inset of Fig. 3(b) of the main manuscript for $U = 0.0075\gamma$ and $\Delta = \gamma$. Reference ~\cite{Casteels16} showed that $\delta I$ exhibits the same double power-law scaling as the hysteresis area. Moreover, in the slow sweep limit the prefactor of the power law with exponent $-1$ is directly related to the tunneling time: $\delta I = 2\tau_{\textrm{tunn}}P_s/t_s$. The above scaling analysis was used to calculate the results in Fig. 3(b) and Fig. 4(b) of the main manuscript.   \\

\noindent\textbf{Comparing measurements and calculations}

A quantitative comparison between experiments and calculations requires a precise knowledge of the experimental excitation efficiency. This is needed to compare driving powers. Since excitation through the cryostat makes it difficult to know the excitation efficiency exactly, we deduced it by requiring that the theoretical range where the transition occurs in Fig. 1(c) of the main manuscript corresponds to the experimental one observed in Fig. 2(d). With this approach, we deduced a conversion factor of $0.75$ between the theoretical units of $\gamma^2$ for the driving intensity and the experimentally measured power  in $\mu W$, i.e.: $I [\gamma^2] =0.75$ Power $[\mu W]$.

The values of $U/\gamma$ used in the calculations in Fig. 3(b) and Fig. 4(b) of the main manuscript were adjusted with respect to the corresponding estimates for the experimental cavities (see above). For the calculations corresponding to cavity 1 and cavity 2, we took the experimentally estimated value of $U/\gamma$  times a factor of  0.4 and  3.5, respectively. Note that the adjusted values are within the experimental uncertainties given above.  These adjustments were only done   to match the measured time of the transition to the power law with exponent -1, and they do not affect in any way the retrieved exponents. For cavity 3, the  estimated  $U/\gamma=2  \substack{+8 \ \\ -1.6 } \cdot 10^{-5}$  was replaced by the value $U/\gamma=0.002$ in the calculations. This modification is due to the extremely small experimental value of $U/\gamma$ being beyond our computational capabilities. Despite this seemingly large adjustment of $U/\gamma$ , both experimental and theoretical curves follow a single power law within the observation window because fluctuations are largely irrelevant to the dynamics given the large average photon number in cavity 3.

In the caption of Fig. 4 we also mention that some of the curves were multiplied by a constant factor (given in the caption) simply for improving the clarity of the figure. These multiplications were only done to improve the visibility of details in the data given the finite space for the figure. These multiplications do not affect in any way the retrieved exponents, nor do they affect the critical times of the transition  from one power law to another; they  only shift the curves down or up in the log scale.

Finally, we would like to mention that a double power law decay of the hysteresis area could also be observed when a nonlinear system is
subjected to thermal (or other) fluctuations. However, the exponents and critical times
can depend on the details of the system. For instance, Luse and Zangwill calculated different exponents for several mean-field treatments of the kinetic Ising model~\cite{Luse94}. In our optical experiments, where laser shot is the only relevant source of
fluctuations, the measured exponents are in  good agreement with  calculations including quantum fluctuations only.


\begin{thebibliography}{55}%
\makeatletter
\providecommand \@ifxundefined [1]{%
 \@ifx{#1\undefined}
}%
\providecommand \@ifnum [1]{%
 \ifnum #1\expandafter \@firstoftwo
 \else \expandafter \@secondoftwo
 \fi
}%
\providecommand \@ifx [1]{%
 \ifx #1\expandafter \@firstoftwo
 \else \expandafter \@secondoftwo
 \fi
}%
\providecommand \natexlab [1]{#1}%
\providecommand \enquote  [1]{``#1''}%
\providecommand \bibnamefont  [1]{#1}%
\providecommand \bibfnamefont [1]{#1}%
\providecommand \citenamefont [1]{#1}%
\providecommand \href@noop [0]{\@secondoftwo}%
\providecommand \href [0]{\begingroup \@sanitize@url \@href}%
\providecommand \@href[1]{\@@startlink{#1}\@@href}%
\providecommand \@@href[1]{\endgroup#1\@@endlink}%
\providecommand \@sanitize@url [0]{\catcode `\\12\catcode `\$12\catcode
  `\&12\catcode `\#12\catcode `\^12\catcode `\_12\catcode `\%12\relax}%
\providecommand \@@startlink[1]{}%
\providecommand \@@endlink[0]{}%
\providecommand \url  [0]{\begingroup\@sanitize@url \@url }%
\providecommand \@url [1]{\endgroup\@href {#1}{\urlprefix }}%
\providecommand \urlprefix  [0]{URL }%
\providecommand \Eprint [0]{\href }%
\providecommand \doibase [0]{http://dx.doi.org/}%
\providecommand \selectlanguage [0]{\@gobble}%
\providecommand \bibinfo  [0]{\@secondoftwo}%
\providecommand \bibfield  [0]{\@secondoftwo}%
\providecommand \translation [1]{[#1]}%
\providecommand \BibitemOpen [0]{}%
\providecommand \bibitemStop [0]{}%
\providecommand \bibitemNoStop [0]{.\EOS\space}%
\providecommand \EOS [0]{\spacefactor3000\relax}%
\providecommand \BibitemShut  [1]{\csname bibitem#1\endcsname}%
\let\auto@bib@innerbib\@empty
\bibitem [{\citenamefont {Gibbs}(1985)}]{Gibbs}%
  \BibitemOpen
  \bibfield  {author} {\bibinfo {author} {\bibfnamefont {H.M.}\ \bibnamefont
  {Gibbs}},\ }\href {https://books.google.fr/books?id=1dVhGQAACAAJ} {\emph
  {\bibinfo {title} {Optical {B}istability: {C}ontrolling {L}ight with
  {L}ight}}},\ Quantum Electronics Series\ (\bibinfo  {publisher} {Academic
  Press},\ \bibinfo {year} {1985})\BibitemShut {NoStop}%
\bibitem [{\citenamefont {Walls}\ and\ \citenamefont
  {Milburn}(2008)}]{WallsMilburn}%
  \BibitemOpen
  \bibfield  {author} {\bibinfo {author} {\bibfnamefont {D.F.}\ \bibnamefont
  {Walls}}\ and\ \bibinfo {author} {\bibfnamefont {G.J.}\ \bibnamefont
  {Milburn}},\ }\href {https://books.google.fr/books?id=LiWsc3Nlf0kC} {\emph
  {\bibinfo {title} {{Q}uantum {O}ptics}}},\ SpringerLink: Springer e-Books\
  (\bibinfo  {publisher} {Springer Berlin Heidelberg},\ \bibinfo {year}
  {2008})\BibitemShut {NoStop}%
\bibitem [{\citenamefont {Bonifacio}\ and\ \citenamefont
  {Lugiato}(1978)}]{Bonifacio78}%
  \BibitemOpen
  \bibfield  {author} {\bibinfo {author} {\bibfnamefont {R.}~\bibnamefont
  {Bonifacio}}\ and\ \bibinfo {author} {\bibfnamefont {L.~A.}\ \bibnamefont
  {Lugiato}},\ }\bibfield  {title} {\enquote {\bibinfo {title} {Photon
  statistics and spectrum of transmitted light in optical bistability},}\
  }\href {\doibase 10.1103/PhysRevLett.40.1023} {\bibfield  {journal} {\bibinfo
   {journal} {Phys. Rev. Lett.}\ }\textbf {\bibinfo {volume} {40}},\ \bibinfo
  {pages} {1023--1027} (\bibinfo {year} {1978})}\BibitemShut {NoStop}%
\bibitem [{\citenamefont {Drummond}\ and\ \citenamefont
  {Walls}(1980)}]{Drummond80}%
  \BibitemOpen
  \bibfield  {author} {\bibinfo {author} {\bibfnamefont {P.~D.}\ \bibnamefont
  {Drummond}}\ and\ \bibinfo {author} {\bibfnamefont {D.~F.}\ \bibnamefont
  {Walls}},\ }\bibfield  {title} {\enquote {\bibinfo {title} {Quantum theory of
  optical bistability. {I}. {N}onlinear polarisability model},}\ }\href
  {http://stacks.iop.org/0305-4470/13/i=2/a=034} {\bibfield  {journal}
  {\bibinfo  {journal} {J. Phys. A}\ }\textbf {\bibinfo {volume} {13}},\
  \bibinfo {pages} {725} (\bibinfo {year} {1980})}\BibitemShut {NoStop}%
\bibitem [{\citenamefont {Roy}\ \emph {et~al.}(1980)\citenamefont {Roy},
  \citenamefont {Short}, \citenamefont {Durnin},\ and\ \citenamefont
  {Mandel}}]{Roy80}%
  \BibitemOpen
  \bibfield  {author} {\bibinfo {author} {\bibfnamefont {Rajarshi}\
  \bibnamefont {Roy}}, \bibinfo {author} {\bibfnamefont {R.}~\bibnamefont
  {Short}}, \bibinfo {author} {\bibfnamefont {J.}~\bibnamefont {Durnin}}, \
  and\ \bibinfo {author} {\bibfnamefont {L.}~\bibnamefont {Mandel}},\
  }\bibfield  {title} {\enquote {\bibinfo {title} {First-passage-time
  distributions under the influence of quantum fluctuations in a laser},}\
  }\href {\doibase 10.1103/PhysRevLett.45.1486} {\bibfield  {journal} {\bibinfo
   {journal} {Phys. Rev. Lett.}\ }\textbf {\bibinfo {volume} {45}},\ \bibinfo
  {pages} {1486--1490} (\bibinfo {year} {1980})}\BibitemShut {NoStop}%
\bibitem [{\citenamefont {Hioe}\ and\ \citenamefont {Singh}(1981)}]{Hioe81}%
  \BibitemOpen
  \bibfield  {author} {\bibinfo {author} {\bibfnamefont {F.~T.}\ \bibnamefont
  {Hioe}}\ and\ \bibinfo {author} {\bibfnamefont {Surendra}\ \bibnamefont
  {Singh}},\ }\bibfield  {title} {\enquote {\bibinfo {title} {Correlations,
  transients, bistability, and phase-transition analogy in two-mode lasers},}\
  }\href {\doibase 10.1103/PhysRevA.24.2050} {\bibfield  {journal} {\bibinfo
  {journal} {Phys. Rev. A}\ }\textbf {\bibinfo {volume} {24}},\ \bibinfo
  {pages} {2050--2074} (\bibinfo {year} {1981})}\BibitemShut {NoStop}%
\bibitem [{\citenamefont {Lett}\ \emph {et~al.}(1981)\citenamefont {Lett},
  \citenamefont {Christian}, \citenamefont {Singh},\ and\ \citenamefont
  {Mandel}}]{Lett81}%
  \BibitemOpen
  \bibfield  {author} {\bibinfo {author} {\bibfnamefont {P.}~\bibnamefont
  {Lett}}, \bibinfo {author} {\bibfnamefont {W.}~\bibnamefont {Christian}},
  \bibinfo {author} {\bibfnamefont {Surendra}\ \bibnamefont {Singh}}, \ and\
  \bibinfo {author} {\bibfnamefont {L.}~\bibnamefont {Mandel}},\ }\bibfield
  {title} {\enquote {\bibinfo {title} {Macroscopic quantum fluctuations and
  first-order phase transition in a laser},}\ }\href {\doibase
  10.1103/PhysRevLett.47.1892} {\bibfield  {journal} {\bibinfo  {journal}
  {Phys. Rev. Lett.}\ }\textbf {\bibinfo {volume} {47}},\ \bibinfo {pages}
  {1892--1895} (\bibinfo {year} {1981})}\BibitemShut {NoStop}%
\bibitem [{\citenamefont {Hartmann}\ \emph {et~al.}(2006)\citenamefont
  {Hartmann}, \citenamefont {Brandao},\ and\ \citenamefont
  {Plenio}}]{Hartmann06}%
  \BibitemOpen
  \bibfield  {author} {\bibinfo {author} {\bibfnamefont {Michael~J}\
  \bibnamefont {Hartmann}}, \bibinfo {author} {\bibfnamefont {Fernando~GSL}\
  \bibnamefont {Brandao}}, \ and\ \bibinfo {author} {\bibfnamefont {Martin~B}\
  \bibnamefont {Plenio}},\ }\bibfield  {title} {\enquote {\bibinfo {title}
  {Strongly interacting polaritons in coupled arrays of cavities},}\ }\href
  {http://www.nature.com/nphys/journal/v2/n12/abs/nphys462.html} {\bibfield
  {journal} {\bibinfo  {journal} {Nature Phys.}\ }\textbf {\bibinfo {volume}
  {2}},\ \bibinfo {pages} {849--855} (\bibinfo {year} {2006})}\BibitemShut
  {NoStop}%
\bibitem [{\citenamefont {Greentree}\ \emph {et~al.}(2006)\citenamefont
  {Greentree}, \citenamefont {Tahan}, \citenamefont {Cole},\ and\ \citenamefont
  {Hollenberg}}]{Greentree06}%
  \BibitemOpen
  \bibfield  {author} {\bibinfo {author} {\bibfnamefont {Andrew~D}\
  \bibnamefont {Greentree}}, \bibinfo {author} {\bibfnamefont {Charles}\
  \bibnamefont {Tahan}}, \bibinfo {author} {\bibfnamefont {Jared~H}\
  \bibnamefont {Cole}}, \ and\ \bibinfo {author} {\bibfnamefont {Lloyd~CL}\
  \bibnamefont {Hollenberg}},\ }\bibfield  {title} {\enquote {\bibinfo {title}
  {Quantum phase transitions of light},}\ }\href
  {http://www.nature.com/nphys/journal/v2/n12/abs/nphys466.html} {\bibfield
  {journal} {\bibinfo  {journal} {Nature Phys.}\ }\textbf {\bibinfo {volume}
  {2}},\ \bibinfo {pages} {856--861} (\bibinfo {year} {2006})}\BibitemShut
  {NoStop}%
\bibitem [{\citenamefont {Angelakis}\ \emph {et~al.}(2007)\citenamefont
  {Angelakis}, \citenamefont {Santos},\ and\ \citenamefont {Bose}}]{Angelakis}%
  \BibitemOpen
  \bibfield  {author} {\bibinfo {author} {\bibfnamefont {Dimitris~G.}\
  \bibnamefont {Angelakis}}, \bibinfo {author} {\bibfnamefont {Marcelo~Franca}\
  \bibnamefont {Santos}}, \ and\ \bibinfo {author} {\bibfnamefont {Sougato}\
  \bibnamefont {Bose}},\ }\bibfield  {title} {\enquote {\bibinfo {title}
  {Photon-blockade-induced {M}ott transitions and {$XY$} spin models in coupled
  cavity arrays},}\ }\href {\doibase 10.1103/PhysRevA.76.031805} {\bibfield
  {journal} {\bibinfo  {journal} {Phys. Rev. A}\ }\textbf {\bibinfo {volume}
  {76}},\ \bibinfo {pages} {031805} (\bibinfo {year} {2007})}\BibitemShut
  {NoStop}%
\bibitem [{\citenamefont {Le~Boit\'e}\ \emph {et~al.}(2013)\citenamefont
  {Le~Boit\'e}, \citenamefont {Orso},\ and\ \citenamefont {Ciuti}}]{LeBoite}%
  \BibitemOpen
  \bibfield  {author} {\bibinfo {author} {\bibfnamefont {Alexandre}\
  \bibnamefont {Le~Boit\'e}}, \bibinfo {author} {\bibfnamefont {Giuliano}\
  \bibnamefont {Orso}}, \ and\ \bibinfo {author} {\bibfnamefont {Cristiano}\
  \bibnamefont {Ciuti}},\ }\bibfield  {title} {\enquote {\bibinfo {title}
  {Steady-state phases and tunneling-induced instabilities in the driven
  dissipative {B}ose-{H}ubbard model},}\ }\href {\doibase
  10.1103/PhysRevLett.110.233601} {\bibfield  {journal} {\bibinfo  {journal}
  {Phys. Rev. Lett.}\ }\textbf {\bibinfo {volume} {110}},\ \bibinfo {pages}
  {233601} (\bibinfo {year} {2013})}\BibitemShut {NoStop}%
\bibitem [{\citenamefont {Wilson}\ \emph {et~al.}(2016)\citenamefont {Wilson},
  \citenamefont {Mahmud}, \citenamefont {Hu}, \citenamefont {Gorshkov},
  \citenamefont {Hafezi},\ and\ \citenamefont {Foss-Feig}}]{Wilson16}%
  \BibitemOpen
  \bibfield  {author} {\bibinfo {author} {\bibfnamefont {Ryan~M.}\ \bibnamefont
  {Wilson}}, \bibinfo {author} {\bibfnamefont {Khan~W.}\ \bibnamefont
  {Mahmud}}, \bibinfo {author} {\bibfnamefont {Anzi}\ \bibnamefont {Hu}},
  \bibinfo {author} {\bibfnamefont {Alexey~V.}\ \bibnamefont {Gorshkov}},
  \bibinfo {author} {\bibfnamefont {Mohammad}\ \bibnamefont {Hafezi}}, \ and\
  \bibinfo {author} {\bibfnamefont {Michael}\ \bibnamefont {Foss-Feig}},\
  }\bibfield  {title} {\enquote {\bibinfo {title} {Collective phases of
  strongly interacting cavity photons},}\ }\href {\doibase
  10.1103/PhysRevA.94.033801} {\bibfield  {journal} {\bibinfo  {journal} {Phys.
  Rev. A}\ }\textbf {\bibinfo {volume} {94}},\ \bibinfo {pages} {033801}
  (\bibinfo {year} {2016})}\BibitemShut {NoStop}%
\bibitem [{\citenamefont {Biondi}\ \emph {et~al.}(2016)\citenamefont {Biondi},
  \citenamefont {Blatter}, \citenamefont {T{\"u}reci},\ and\ \citenamefont
  {Schmidt}}]{Biondi16}%
  \BibitemOpen
  \bibfield  {author} {\bibinfo {author} {\bibfnamefont {Matteo}\ \bibnamefont
  {Biondi}}, \bibinfo {author} {\bibfnamefont {Gianni}\ \bibnamefont
  {Blatter}}, \bibinfo {author} {\bibfnamefont {Hakan~E}\ \bibnamefont
  {T{\"u}reci}}, \ and\ \bibinfo {author} {\bibfnamefont {Sebastian}\
  \bibnamefont {Schmidt}},\ }\bibfield  {title} {\enquote {\bibinfo {title}
  {Nonequilibrium phase diagram of the driven-dissipative photonic lattice},}\
  }\href {https://arxiv.org/abs/1611.00697} {\bibfield  {journal} {\bibinfo
  {journal} {arXiv:1611.00697}\ } (\bibinfo {year} {2016})}\BibitemShut
  {NoStop}%
\bibitem [{\citenamefont {Carmichael}(2015)}]{Carmichael15}%
  \BibitemOpen
  \bibfield  {author} {\bibinfo {author} {\bibfnamefont {H.~J.}\ \bibnamefont
  {Carmichael}},\ }\bibfield  {title} {\enquote {\bibinfo {title} {Breakdown of
  photon blockade: A dissipative quantum phase transition in zero
  dimensions},}\ }\href {\doibase 10.1103/PhysRevX.5.031028} {\bibfield
  {journal} {\bibinfo  {journal} {Phys. Rev. X}\ }\textbf {\bibinfo {volume}
  {5}},\ \bibinfo {pages} {031028} (\bibinfo {year} {2015})}\BibitemShut
  {NoStop}%
\bibitem [{\citenamefont {Casteels}\ \emph
  {et~al.}(2016{\natexlab{a}})\citenamefont {Casteels}, \citenamefont {Storme},
  \citenamefont {Le~Boit\'e},\ and\ \citenamefont {Ciuti}}]{Casteels16}%
  \BibitemOpen
  \bibfield  {author} {\bibinfo {author} {\bibfnamefont {W.}~\bibnamefont
  {Casteels}}, \bibinfo {author} {\bibfnamefont {F.}~\bibnamefont {Storme}},
  \bibinfo {author} {\bibfnamefont {A.}~\bibnamefont {Le~Boit\'e}}, \ and\
  \bibinfo {author} {\bibfnamefont {C.}~\bibnamefont {Ciuti}},\ }\bibfield
  {title} {\enquote {\bibinfo {title} {Power laws in the dynamic hysteresis of
  quantum nonlinear photonic resonators},}\ }\href {\doibase
  10.1103/PhysRevA.93.033824} {\bibfield  {journal} {\bibinfo  {journal} {Phys.
  Rev. A}\ }\textbf {\bibinfo {volume} {93}},\ \bibinfo {pages} {033824}
  (\bibinfo {year} {2016}{\natexlab{a}})}\BibitemShut {NoStop}%
\bibitem [{\citenamefont {Mendoza-Arenas}\ \emph {et~al.}(2016)\citenamefont
  {Mendoza-Arenas}, \citenamefont {Clark}, \citenamefont {Felicetti},
  \citenamefont {Romero}, \citenamefont {Solano}, \citenamefont {Angelakis},\
  and\ \citenamefont {Jaksch}}]{Mendoza16}%
  \BibitemOpen
  \bibfield  {author} {\bibinfo {author} {\bibfnamefont {J.~J.}\ \bibnamefont
  {Mendoza-Arenas}}, \bibinfo {author} {\bibfnamefont {S.~R.}\ \bibnamefont
  {Clark}}, \bibinfo {author} {\bibfnamefont {S.}~\bibnamefont {Felicetti}},
  \bibinfo {author} {\bibfnamefont {G.}~\bibnamefont {Romero}}, \bibinfo
  {author} {\bibfnamefont {E.}~\bibnamefont {Solano}}, \bibinfo {author}
  {\bibfnamefont {D.~G.}\ \bibnamefont {Angelakis}}, \ and\ \bibinfo {author}
  {\bibfnamefont {D.}~\bibnamefont {Jaksch}},\ }\bibfield  {title} {\enquote
  {\bibinfo {title} {Beyond mean-field bistability in driven-dissipative
  lattices: Bunching-antibunching transition and quantum simulation},}\ }\href
  {\doibase 10.1103/PhysRevA.93.023821} {\bibfield  {journal} {\bibinfo
  {journal} {Phys. Rev. A}\ }\textbf {\bibinfo {volume} {93}},\ \bibinfo
  {pages} {023821} (\bibinfo {year} {2016})}\BibitemShut {NoStop}%
\bibitem [{\citenamefont {Casteels}\ \emph
  {et~al.}(2016{\natexlab{b}})\citenamefont {Casteels}, \citenamefont {Fazio},\
  and\ \citenamefont {Ciuti}}]{Casteels16arxiv}%
  \BibitemOpen
  \bibfield  {author} {\bibinfo {author} {\bibfnamefont {Wim}\ \bibnamefont
  {Casteels}}, \bibinfo {author} {\bibfnamefont {Rosario}\ \bibnamefont
  {Fazio}}, \ and\ \bibinfo {author} {\bibfnamefont {Cristiano}\ \bibnamefont
  {Ciuti}},\ }\bibfield  {title} {\enquote {\bibinfo {title} {Critical scaling
  of the liouvillian gap for a nonlinear driven-dissipative resonator},}\
  }\href {https://arxiv.org/abs/1608.00717} {\bibfield  {journal} {\bibinfo
  {journal} {arXiv:1608.00717}\ } (\bibinfo {year}
  {2016}{\natexlab{b}})}\BibitemShut {NoStop}%
\bibitem [{\citenamefont {Foss-Feig}\ \emph {et~al.}(2016)\citenamefont
  {Foss-Feig}, \citenamefont {Niroula}, \citenamefont {Young}, \citenamefont
  {Hafezi}, \citenamefont {Gorshkov}, \citenamefont {Wilson},\ and\
  \citenamefont {Maghrebi}}]{Foss16}%
  \BibitemOpen
  \bibfield  {author} {\bibinfo {author} {\bibfnamefont {Michael}\ \bibnamefont
  {Foss-Feig}}, \bibinfo {author} {\bibfnamefont {Pradeep}\ \bibnamefont
  {Niroula}}, \bibinfo {author} {\bibfnamefont {Jeremy~T}\ \bibnamefont
  {Young}}, \bibinfo {author} {\bibfnamefont {Mohammad}\ \bibnamefont
  {Hafezi}}, \bibinfo {author} {\bibfnamefont {Alexey~V}\ \bibnamefont
  {Gorshkov}}, \bibinfo {author} {\bibfnamefont {Ryan~M}\ \bibnamefont
  {Wilson}}, \ and\ \bibinfo {author} {\bibfnamefont {Mohammad~F}\ \bibnamefont
  {Maghrebi}},\ }\bibfield  {title} {\enquote {\bibinfo {title} {Emergent
  equilibrium in many-body optical bistability},}\ }\href
  {https://arxiv.org/abs/1611.02284} {\bibfield  {journal} {\bibinfo  {journal}
  {arXiv:1611.02284}\ } (\bibinfo {year} {2016})}\BibitemShut {NoStop}%
\bibitem [{\citenamefont {Kessler}\ \emph {et~al.}(2012)\citenamefont
  {Kessler}, \citenamefont {Giedke}, \citenamefont {Imamoglu}, \citenamefont
  {Yelin}, \citenamefont {Lukin},\ and\ \citenamefont {Cirac}}]{Kessler12}%
  \BibitemOpen
  \bibfield  {author} {\bibinfo {author} {\bibfnamefont {E.~M.}\ \bibnamefont
  {Kessler}}, \bibinfo {author} {\bibfnamefont {G.}~\bibnamefont {Giedke}},
  \bibinfo {author} {\bibfnamefont {A.}~\bibnamefont {Imamoglu}}, \bibinfo
  {author} {\bibfnamefont {S.~F.}\ \bibnamefont {Yelin}}, \bibinfo {author}
  {\bibfnamefont {M.~D.}\ \bibnamefont {Lukin}}, \ and\ \bibinfo {author}
  {\bibfnamefont {J.~I.}\ \bibnamefont {Cirac}},\ }\bibfield  {title} {\enquote
  {\bibinfo {title} {Dissipative phase transition in a central spin system},}\
  }\href {\doibase 10.1103/PhysRevA.86.012116} {\bibfield  {journal} {\bibinfo
  {journal} {Phys. Rev. A}\ }\textbf {\bibinfo {volume} {86}},\ \bibinfo
  {pages} {012116} (\bibinfo {year} {2012})}\BibitemShut {NoStop}%
\bibitem [{\citenamefont {Majumdar}\ \emph {et~al.}(2012)\citenamefont
  {Majumdar}, \citenamefont {Rundquist}, \citenamefont {Bajcsy}, \citenamefont
  {Dasika}, \citenamefont {Bank},\ and\ \citenamefont {Vu\ifmmode
  \check{c}\else \v{c}\fi{}kovi\ifmmode~\acute{c}\else \'{c}\fi{}}}]{Vuck12}%
  \BibitemOpen
  \bibfield  {author} {\bibinfo {author} {\bibfnamefont {Arka}\ \bibnamefont
  {Majumdar}}, \bibinfo {author} {\bibfnamefont {Armand}\ \bibnamefont
  {Rundquist}}, \bibinfo {author} {\bibfnamefont {Michal}\ \bibnamefont
  {Bajcsy}}, \bibinfo {author} {\bibfnamefont {Vaishno~D.}\ \bibnamefont
  {Dasika}}, \bibinfo {author} {\bibfnamefont {Seth~R.}\ \bibnamefont {Bank}},
  \ and\ \bibinfo {author} {\bibfnamefont {Jelena}\ \bibnamefont {Vu\ifmmode
  \check{c}\else \v{c}\fi{}kovi\ifmmode~\acute{c}\else \'{c}\fi{}}},\
  }\bibfield  {title} {\enquote {\bibinfo {title} {Design and analysis of
  photonic crystal coupled cavity arrays for quantum simulation},}\ }\href
  {\doibase 10.1103/PhysRevB.86.195312} {\bibfield  {journal} {\bibinfo
  {journal} {Phys. Rev. B}\ }\textbf {\bibinfo {volume} {86}},\ \bibinfo
  {pages} {195312} (\bibinfo {year} {2012})}\BibitemShut {NoStop}%
\bibitem [{\citenamefont {Hamel}\ \emph {et~al.}(2015)\citenamefont {Hamel},
  \citenamefont {Haddadi}, \citenamefont {Raineri}, \citenamefont {Monnier},
  \citenamefont {Beaudoin}, \citenamefont {Sagnes}, \citenamefont {Levenson},\
  and\ \citenamefont {Yacomotti}}]{Hamel15}%
  \BibitemOpen
  \bibfield  {author} {\bibinfo {author} {\bibfnamefont {Philippe}\
  \bibnamefont {Hamel}}, \bibinfo {author} {\bibfnamefont {Samir}\ \bibnamefont
  {Haddadi}}, \bibinfo {author} {\bibfnamefont {Fabrice}\ \bibnamefont
  {Raineri}}, \bibinfo {author} {\bibfnamefont {Paul}\ \bibnamefont {Monnier}},
  \bibinfo {author} {\bibfnamefont {Gregoire}\ \bibnamefont {Beaudoin}},
  \bibinfo {author} {\bibfnamefont {Isabelle}\ \bibnamefont {Sagnes}}, \bibinfo
  {author} {\bibfnamefont {Ariel}\ \bibnamefont {Levenson}}, \ and\ \bibinfo
  {author} {\bibfnamefont {Alejandro~M}\ \bibnamefont {Yacomotti}},\ }\bibfield
   {title} {\enquote {\bibinfo {title} {Spontaneous mirror-symmetry breaking in
  coupled photonic-crystal nanolasers},}\ }\href
  {http://www.nature.com/nphoton/journal/v9/n5/full/nphoton.2015.65.html}
  {\bibfield  {journal} {\bibinfo  {journal} {Nature Photon.}\ }\textbf
  {\bibinfo {volume} {9}},\ \bibinfo {pages} {311--315} (\bibinfo {year}
  {2015})}\BibitemShut {NoStop}%
\bibitem [{\citenamefont {Fleischer}\ \emph {et~al.}(2003)\citenamefont
  {Fleischer}, \citenamefont {Segev}, \citenamefont {Efremidis},\ and\
  \citenamefont {Christodoulides}}]{Fleischer03}%
  \BibitemOpen
  \bibfield  {author} {\bibinfo {author} {\bibfnamefont {Jason~W}\ \bibnamefont
  {Fleischer}}, \bibinfo {author} {\bibfnamefont {Mordechai}\ \bibnamefont
  {Segev}}, \bibinfo {author} {\bibfnamefont {Nikolaos~K}\ \bibnamefont
  {Efremidis}}, \ and\ \bibinfo {author} {\bibfnamefont {Demetrios~N}\
  \bibnamefont {Christodoulides}},\ }\bibfield  {title} {\enquote {\bibinfo
  {title} {Observation of two-dimensional discrete solitons in optically
  induced nonlinear photonic lattices},}\ }\href
  {http://www.nature.com/nature/journal/v422/n6928/full/nature01452.html}
  {\bibfield  {journal} {\bibinfo  {journal} {Nature}\ }\textbf {\bibinfo
  {volume} {422}},\ \bibinfo {pages} {147--150} (\bibinfo {year}
  {2003})}\BibitemShut {NoStop}%
\bibitem [{\citenamefont {Vijay}\ \emph {et~al.}(2009)\citenamefont {Vijay},
  \citenamefont {Devoret},\ and\ \citenamefont {Siddiqi}}]{Vijay09}%
  \BibitemOpen
  \bibfield  {author} {\bibinfo {author} {\bibfnamefont {R.}~\bibnamefont
  {Vijay}}, \bibinfo {author} {\bibfnamefont {M.~H.}\ \bibnamefont {Devoret}},
  \ and\ \bibinfo {author} {\bibfnamefont {I.}~\bibnamefont {Siddiqi}},\
  }\bibfield  {title} {\enquote {\bibinfo {title} {Invited review article: The
  {J}osephson bifurcation amplifier},}\ }\href
  {http://scitation.aip.org/content/aip/journal/rsi/80/11/10.1063/1.3224703}
  {\bibfield  {journal} {\bibinfo  {journal} {Rev. Sci. Instrum.}\ }\textbf
  {\bibinfo {volume} {80}},\ \bibinfo {eid} {111101} (\bibinfo {year}
  {2009})}\BibitemShut {NoStop}%
\bibitem [{\citenamefont {Underwood}\ \emph {et~al.}(2012)\citenamefont
  {Underwood}, \citenamefont {Shanks}, \citenamefont {Koch},\ and\
  \citenamefont {Houck}}]{Underwood12}%
  \BibitemOpen
  \bibfield  {author} {\bibinfo {author} {\bibfnamefont {D.~L.}\ \bibnamefont
  {Underwood}}, \bibinfo {author} {\bibfnamefont {W.~E.}\ \bibnamefont
  {Shanks}}, \bibinfo {author} {\bibfnamefont {Jens}\ \bibnamefont {Koch}}, \
  and\ \bibinfo {author} {\bibfnamefont {A.~A.}\ \bibnamefont {Houck}},\
  }\bibfield  {title} {\enquote {\bibinfo {title} {Low-disorder microwave
  cavity lattices for quantum simulation with photons},}\ }\href {\doibase
  10.1103/PhysRevA.86.023837} {\bibfield  {journal} {\bibinfo  {journal} {Phys.
  Rev. A}\ }\textbf {\bibinfo {volume} {86}},\ \bibinfo {pages} {023837}
  (\bibinfo {year} {2012})}\BibitemShut {NoStop}%
\bibitem [{\citenamefont {Eichenfield}\ \emph {et~al.}(2009)\citenamefont
  {Eichenfield}, \citenamefont {Chan}, \citenamefont {Camacho}, \citenamefont
  {Vahala},\ and\ \citenamefont {Painter}}]{Painter09}%
  \BibitemOpen
  \bibfield  {author} {\bibinfo {author} {\bibfnamefont {Matt}\ \bibnamefont
  {Eichenfield}}, \bibinfo {author} {\bibfnamefont {Jasper}\ \bibnamefont
  {Chan}}, \bibinfo {author} {\bibfnamefont {Ryan~M}\ \bibnamefont {Camacho}},
  \bibinfo {author} {\bibfnamefont {Kerry~J}\ \bibnamefont {Vahala}}, \ and\
  \bibinfo {author} {\bibfnamefont {Oskar}\ \bibnamefont {Painter}},\
  }\bibfield  {title} {\enquote {\bibinfo {title} {Optomechanical crystals},}\
  }\href {http://www.nature.com/nature/journal/v462/n7269/abs/nature08524.html}
  {\bibfield  {journal} {\bibinfo  {journal} {Nature}\ }\textbf {\bibinfo
  {volume} {462}},\ \bibinfo {pages} {78--82} (\bibinfo {year}
  {2009})}\BibitemShut {NoStop}%
\bibitem [{\citenamefont {Gil-Santos}\ \emph {et~al.}(2016)\citenamefont
  {Gil-Santos}, \citenamefont {Labousse}, \citenamefont {Baker}, \citenamefont
  {Goetschy}, \citenamefont {Hease}, \citenamefont {Gomez}, \citenamefont
  {Lema{\^\i}tre}, \citenamefont {Leo}, \citenamefont {Ciuti},\ and\
  \citenamefont {Favero}}]{Favero16}%
  \BibitemOpen
  \bibfield  {author} {\bibinfo {author} {\bibfnamefont {Eduardo}\ \bibnamefont
  {Gil-Santos}}, \bibinfo {author} {\bibfnamefont {Matthieu}\ \bibnamefont
  {Labousse}}, \bibinfo {author} {\bibfnamefont {Christophe}\ \bibnamefont
  {Baker}}, \bibinfo {author} {\bibfnamefont {Arthur}\ \bibnamefont
  {Goetschy}}, \bibinfo {author} {\bibfnamefont {William}\ \bibnamefont
  {Hease}}, \bibinfo {author} {\bibfnamefont {Carmen}\ \bibnamefont {Gomez}},
  \bibinfo {author} {\bibfnamefont {Aristide}\ \bibnamefont {Lema{\^\i}tre}},
  \bibinfo {author} {\bibfnamefont {Giuseppe}\ \bibnamefont {Leo}}, \bibinfo
  {author} {\bibfnamefont {Cristiano}\ \bibnamefont {Ciuti}}, \ and\ \bibinfo
  {author} {\bibfnamefont {Ivan}\ \bibnamefont {Favero}},\ }\bibfield  {title}
  {\enquote {\bibinfo {title} {Light-mediated cascaded locking of multiple
  nano-optomechanical oscillators},}\ }\href {https://arxiv.org/abs/1609.09712}
  {\bibfield  {journal} {\bibinfo  {journal} {arXiv:1609.09712}\ } (\bibinfo
  {year} {2016})}\BibitemShut {NoStop}%
\bibitem [{\citenamefont {Carusotto}\ and\ \citenamefont
  {Ciuti}(2013)}]{CarusottoRMP}%
  \BibitemOpen
  \bibfield  {author} {\bibinfo {author} {\bibfnamefont {Iacopo}\ \bibnamefont
  {Carusotto}}\ and\ \bibinfo {author} {\bibfnamefont {Cristiano}\ \bibnamefont
  {Ciuti}},\ }\bibfield  {title} {\enquote {\bibinfo {title} {Quantum fluids of
  light},}\ }\href {\doibase 10.1103/RevModPhys.85.299} {\bibfield  {journal}
  {\bibinfo  {journal} {Rev. Mod. Phys.}\ }\textbf {\bibinfo {volume} {85}},\
  \bibinfo {pages} {299--366} (\bibinfo {year} {2013})}\BibitemShut {NoStop}%
\bibitem [{\citenamefont {Kim}\ \emph {et~al.}(2011)\citenamefont {Kim},
  \citenamefont {Kusudo}, \citenamefont {Wu}, \citenamefont {Masumoto},
  \citenamefont {L{\"o}ffler}, \citenamefont {H{\"o}fling}, \citenamefont
  {Kumada}, \citenamefont {Worschech}, \citenamefont {Forchel},\ and\
  \citenamefont {Yamamoto}}]{Kim11}%
  \BibitemOpen
  \bibfield  {author} {\bibinfo {author} {\bibfnamefont {Na~Young}\
  \bibnamefont {Kim}}, \bibinfo {author} {\bibfnamefont {Kenichiro}\
  \bibnamefont {Kusudo}}, \bibinfo {author} {\bibfnamefont {Congjun}\
  \bibnamefont {Wu}}, \bibinfo {author} {\bibfnamefont {Naoyuki}\ \bibnamefont
  {Masumoto}}, \bibinfo {author} {\bibfnamefont {Andreas}\ \bibnamefont
  {L{\"o}ffler}}, \bibinfo {author} {\bibfnamefont {Sven}\ \bibnamefont
  {H{\"o}fling}}, \bibinfo {author} {\bibfnamefont {Norio}\ \bibnamefont
  {Kumada}}, \bibinfo {author} {\bibfnamefont {Lukas}\ \bibnamefont
  {Worschech}}, \bibinfo {author} {\bibfnamefont {Alfred}\ \bibnamefont
  {Forchel}}, \ and\ \bibinfo {author} {\bibfnamefont {Yoshihisa}\ \bibnamefont
  {Yamamoto}},\ }\bibfield  {title} {\enquote {\bibinfo {title} {Dynamical
  d-wave condensation of exciton-polaritons in a two-dimensional square-lattice
  potential},}\ }\href
  {http://www.nature.com/nphys/journal/v7/n9/full/nphys2012.html} {\bibfield
  {journal} {\bibinfo  {journal} {Nat. Phys.}\ }\textbf {\bibinfo {volume}
  {7}},\ \bibinfo {pages} {681--686} (\bibinfo {year} {2011})}\BibitemShut
  {NoStop}%
\bibitem [{\citenamefont {Baboux}\ \emph {et~al.}(2016)\citenamefont {Baboux},
  \citenamefont {Ge}, \citenamefont {Jacqmin}, \citenamefont {Biondi},
  \citenamefont {Galopin}, \citenamefont {Lema\^{\i}tre}, \citenamefont
  {Le~Gratiet}, \citenamefont {Sagnes}, \citenamefont {Schmidt}, \citenamefont
  {T\"ureci}, \citenamefont {Amo},\ and\ \citenamefont {Bloch}}]{Baboux16}%
  \BibitemOpen
  \bibfield  {author} {\bibinfo {author} {\bibfnamefont {F.}~\bibnamefont
  {Baboux}}, \bibinfo {author} {\bibfnamefont {L.}~\bibnamefont {Ge}}, \bibinfo
  {author} {\bibfnamefont {T.}~\bibnamefont {Jacqmin}}, \bibinfo {author}
  {\bibfnamefont {M.}~\bibnamefont {Biondi}}, \bibinfo {author} {\bibfnamefont
  {E.}~\bibnamefont {Galopin}}, \bibinfo {author} {\bibfnamefont
  {A.}~\bibnamefont {Lema\^{\i}tre}}, \bibinfo {author} {\bibfnamefont
  {L.}~\bibnamefont {Le~Gratiet}}, \bibinfo {author} {\bibfnamefont
  {I.}~\bibnamefont {Sagnes}}, \bibinfo {author} {\bibfnamefont
  {S.}~\bibnamefont {Schmidt}}, \bibinfo {author} {\bibfnamefont {H.~E.}\
  \bibnamefont {T\"ureci}}, \bibinfo {author} {\bibfnamefont {A.}~\bibnamefont
  {Amo}}, \ and\ \bibinfo {author} {\bibfnamefont {J.}~\bibnamefont {Bloch}},\
  }\bibfield  {title} {\enquote {\bibinfo {title} {Bosonic condensation and
  disorder-induced localization in a flat band},}\ }\href {\doibase
  10.1103/PhysRevLett.116.066402} {\bibfield  {journal} {\bibinfo  {journal}
  {Phys. Rev. Lett.}\ }\textbf {\bibinfo {volume} {116}},\ \bibinfo {pages}
  {066402} (\bibinfo {year} {2016})}\BibitemShut {NoStop}%
\bibitem [{\citenamefont {Amo}\ \emph {et~al.}(2009)\citenamefont {Amo} \emph
  {et~al.}}]{Amo09}%
  \BibitemOpen
  \bibfield  {author} {\bibinfo {author} {\bibfnamefont {Alberto}\ \bibnamefont
  {Amo}} \emph {et~al.},\ }\bibfield  {title} {\enquote {\bibinfo {title}
  {Superfluidity of polaritons in semiconductor microcavities},}\ }\href
  {http://www.nature.com/nphys/journal/v5/n11/abs/nphys1364.html} {\bibfield
  {journal} {\bibinfo  {journal} {Nat. Phys.}\ }\textbf {\bibinfo {volume}
  {5}},\ \bibinfo {pages} {805} (\bibinfo {year} {2009})}\BibitemShut {NoStop}%
\bibitem [{\citenamefont {Baas}\ \emph {et~al.}(2004)\citenamefont {Baas},
  \citenamefont {Karr}, \citenamefont {Eleuch},\ and\ \citenamefont
  {Giacobino}}]{Baas04}%
  \BibitemOpen
  \bibfield  {author} {\bibinfo {author} {\bibfnamefont {A.}~\bibnamefont
  {Baas}}, \bibinfo {author} {\bibfnamefont {J.~Ph.}\ \bibnamefont {Karr}},
  \bibinfo {author} {\bibfnamefont {H.}~\bibnamefont {Eleuch}}, \ and\ \bibinfo
  {author} {\bibfnamefont {E.}~\bibnamefont {Giacobino}},\ }\bibfield  {title}
  {\enquote {\bibinfo {title} {Optical bistability in semiconductor
  microcavities},}\ }\href {\doibase 10.1103/PhysRevA.69.023809} {\bibfield
  {journal} {\bibinfo  {journal} {Phys. Rev. A}\ }\textbf {\bibinfo {volume}
  {69}},\ \bibinfo {pages} {023809} (\bibinfo {year} {2004})}\BibitemShut
  {NoStop}%
\bibitem [{\citenamefont {Para{\"\i}so}\ \emph {et~al.}(2010)\citenamefont
  {Para{\"\i}so}, \citenamefont {Wouters}, \citenamefont {L{\'e}ger},
  \citenamefont {Morier-Genoud},\ and\ \citenamefont
  {Deveaud-Pl{\'e}dran}}]{Paraiso10}%
  \BibitemOpen
  \bibfield  {author} {\bibinfo {author} {\bibfnamefont {TK}~\bibnamefont
  {Para{\"\i}so}}, \bibinfo {author} {\bibfnamefont {M}~\bibnamefont
  {Wouters}}, \bibinfo {author} {\bibfnamefont {Y}~\bibnamefont {L{\'e}ger}},
  \bibinfo {author} {\bibfnamefont {F}~\bibnamefont {Morier-Genoud}}, \ and\
  \bibinfo {author} {\bibfnamefont {B}~\bibnamefont {Deveaud-Pl{\'e}dran}},\
  }\bibfield  {title} {\enquote {\bibinfo {title} {Multistability of a coherent
  spin ensemble in a semiconductor microcavity},}\ }\href
  {http://www.nature.com/nmat/journal/v9/n8/abs/nmat2787.html} {\bibfield
  {journal} {\bibinfo  {journal} {Nature {M}ater.}\ }\textbf {\bibinfo {volume}
  {9}},\ \bibinfo {pages} {655--660} (\bibinfo {year} {2010})}\BibitemShut
  {NoStop}%
\bibitem [{\citenamefont {Rodriguez}\ \emph {et~al.}(2016)\citenamefont
  {Rodriguez}, \citenamefont {Amo}, \citenamefont {Sagnes}, \citenamefont
  {Le~Gratiet}, \citenamefont {Galopin}, \citenamefont {Lemaitre},\ and\
  \citenamefont {Bloch}}]{Rodriguez16}%
  \BibitemOpen
  \bibfield  {author} {\bibinfo {author} {\bibfnamefont {SRK}\ \bibnamefont
  {Rodriguez}}, \bibinfo {author} {\bibfnamefont {A}~\bibnamefont {Amo}},
  \bibinfo {author} {\bibfnamefont {I}~\bibnamefont {Sagnes}}, \bibinfo
  {author} {\bibfnamefont {L}~\bibnamefont {Le~Gratiet}}, \bibinfo {author}
  {\bibfnamefont {E}~\bibnamefont {Galopin}}, \bibinfo {author} {\bibfnamefont
  {A}~\bibnamefont {Lemaitre}}, \ and\ \bibinfo {author} {\bibfnamefont
  {J}~\bibnamefont {Bloch}},\ }\bibfield  {title} {\enquote {\bibinfo {title}
  {Interaction-induced hopping phase in driven-dissipative coupled photonic
  microcavities},}\ }\href
  {http://www.nature.com/ncomms/2016/160616/ncomms11887/full/ncomms11887.html}
  {\bibfield  {journal} {\bibinfo  {journal} {Nature Commun.}\ }\textbf
  {\bibinfo {volume} {7}},\ \bibinfo {pages} {11887} (\bibinfo {year}
  {2016})}\BibitemShut {NoStop}%
\bibitem [{\citenamefont {Luse}\ and\ \citenamefont {Zangwill}(1994)}]{Luse94}%
  \BibitemOpen
  \bibfield  {author} {\bibinfo {author} {\bibfnamefont {C.~N.}\ \bibnamefont
  {Luse}}\ and\ \bibinfo {author} {\bibfnamefont {A.}~\bibnamefont
  {Zangwill}},\ }\bibfield  {title} {\enquote {\bibinfo {title} {Discontinuous
  scaling of hysteresis losses},}\ }\href {\doibase 10.1103/PhysRevE.50.224}
  {\bibfield  {journal} {\bibinfo  {journal} {Phys. Rev. E}\ }\textbf {\bibinfo
  {volume} {50}},\ \bibinfo {pages} {224--226} (\bibinfo {year}
  {1994})}\BibitemShut {NoStop}%
\bibitem [{\citenamefont {Kerckhoff}\ \emph {et~al.}(2011)\citenamefont
  {Kerckhoff}, \citenamefont {Armen},\ and\ \citenamefont
  {Mabuchi}}]{Mabuchi11}%
  \BibitemOpen
  \bibfield  {author} {\bibinfo {author} {\bibfnamefont {Joseph}\ \bibnamefont
  {Kerckhoff}}, \bibinfo {author} {\bibfnamefont {Michael~A.}\ \bibnamefont
  {Armen}}, \ and\ \bibinfo {author} {\bibfnamefont {Hideo}\ \bibnamefont
  {Mabuchi}},\ }\bibfield  {title} {\enquote {\bibinfo {title} {Remnants of
  semiclassical bistability in the few-photon regime of cavity {QED}},}\ }\href
  {\doibase 10.1364/OE.19.024468} {\bibfield  {journal} {\bibinfo  {journal}
  {Opt. Express}\ }\textbf {\bibinfo {volume} {19}},\ \bibinfo {pages}
  {24468--24482} (\bibinfo {year} {2011})}\BibitemShut {NoStop}%
\bibitem [{\citenamefont {Abbaspour}\ \emph {et~al.}(2015)\citenamefont
  {Abbaspour}, \citenamefont {Sallen}, \citenamefont {Trebaol}, \citenamefont
  {Morier-Genoud}, \citenamefont {Portella-Oberli},\ and\ \citenamefont
  {Deveaud}}]{Deveaud15}%
  \BibitemOpen
  \bibfield  {author} {\bibinfo {author} {\bibfnamefont {H.}~\bibnamefont
  {Abbaspour}}, \bibinfo {author} {\bibfnamefont {G.}~\bibnamefont {Sallen}},
  \bibinfo {author} {\bibfnamefont {S.}~\bibnamefont {Trebaol}}, \bibinfo
  {author} {\bibfnamefont {F.}~\bibnamefont {Morier-Genoud}}, \bibinfo {author}
  {\bibfnamefont {M.~T.}\ \bibnamefont {Portella-Oberli}}, \ and\ \bibinfo
  {author} {\bibfnamefont {B.}~\bibnamefont {Deveaud}},\ }\bibfield  {title}
  {\enquote {\bibinfo {title} {Effect of a noisy driving field on a bistable
  polariton system},}\ }\href {\doibase 10.1103/PhysRevB.92.165303} {\bibfield
  {journal} {\bibinfo  {journal} {Phys. Rev. B}\ }\textbf {\bibinfo {volume}
  {92}},\ \bibinfo {pages} {165303} (\bibinfo {year} {2015})}\BibitemShut
  {NoStop}%
\bibitem [{\citenamefont {Gibbs}\ \emph {et~al.}(1976)\citenamefont {Gibbs},
  \citenamefont {McCall},\ and\ \citenamefont {Venkatesan}}]{Gibbs76}%
  \BibitemOpen
  \bibfield  {author} {\bibinfo {author} {\bibfnamefont {H.~M.}\ \bibnamefont
  {Gibbs}}, \bibinfo {author} {\bibfnamefont {S.~L.}\ \bibnamefont {McCall}}, \
  and\ \bibinfo {author} {\bibfnamefont {T.~N.~C.}\ \bibnamefont
  {Venkatesan}},\ }\bibfield  {title} {\enquote {\bibinfo {title} {Differential
  gain and bistability using a sodium-filled {F}abry-{P}erot interferometer},}\
  }\href {\doibase 10.1103/PhysRevLett.36.1135} {\bibfield  {journal} {\bibinfo
   {journal} {Phys. Rev. Lett.}\ }\textbf {\bibinfo {volume} {36}},\ \bibinfo
  {pages} {1135--1138} (\bibinfo {year} {1976})}\BibitemShut {NoStop}%
\bibitem [{\citenamefont {Dorsel}\ \emph {et~al.}(1983)\citenamefont {Dorsel},
  \citenamefont {McCullen}, \citenamefont {Meystre}, \citenamefont {Vignes},\
  and\ \citenamefont {Walther}}]{Dorsel83}%
  \BibitemOpen
  \bibfield  {author} {\bibinfo {author} {\bibfnamefont {A.}~\bibnamefont
  {Dorsel}}, \bibinfo {author} {\bibfnamefont {J.~D.}\ \bibnamefont
  {McCullen}}, \bibinfo {author} {\bibfnamefont {P.}~\bibnamefont {Meystre}},
  \bibinfo {author} {\bibfnamefont {E.}~\bibnamefont {Vignes}}, \ and\ \bibinfo
  {author} {\bibfnamefont {H.}~\bibnamefont {Walther}},\ }\bibfield  {title}
  {\enquote {\bibinfo {title} {Optical bistability and mirror confinement
  induced by radiation pressure},}\ }\href {\doibase
  10.1103/PhysRevLett.51.1550} {\bibfield  {journal} {\bibinfo  {journal}
  {Phys. Rev. Lett.}\ }\textbf {\bibinfo {volume} {51}},\ \bibinfo {pages}
  {1550--1553} (\bibinfo {year} {1983})}\BibitemShut {NoStop}%
\bibitem [{\citenamefont {Jung}\ \emph {et~al.}(1990)\citenamefont {Jung},
  \citenamefont {Gray}, \citenamefont {Roy},\ and\ \citenamefont
  {Mandel}}]{Mandel90}%
  \BibitemOpen
  \bibfield  {author} {\bibinfo {author} {\bibfnamefont {Peter}\ \bibnamefont
  {Jung}}, \bibinfo {author} {\bibfnamefont {George}\ \bibnamefont {Gray}},
  \bibinfo {author} {\bibfnamefont {Rajarshi}\ \bibnamefont {Roy}}, \ and\
  \bibinfo {author} {\bibfnamefont {Paul}\ \bibnamefont {Mandel}},\ }\bibfield
  {title} {\enquote {\bibinfo {title} {Scaling law for dynamical hysteresis},}\
  }\href {\doibase 10.1103/PhysRevLett.65.1873} {\bibfield  {journal} {\bibinfo
   {journal} {Phys. Rev. Lett.}\ }\textbf {\bibinfo {volume} {65}},\ \bibinfo
  {pages} {1873--1876} (\bibinfo {year} {1990})}\BibitemShut {NoStop}%
\bibitem [{\citenamefont {Rempe}\ \emph {et~al.}(1991)\citenamefont {Rempe},
  \citenamefont {Thompson}, \citenamefont {Brecha}, \citenamefont {Lee},\ and\
  \citenamefont {Kimble}}]{Rempe91}%
  \BibitemOpen
  \bibfield  {author} {\bibinfo {author} {\bibfnamefont {G.}~\bibnamefont
  {Rempe}}, \bibinfo {author} {\bibfnamefont {R.~J.}\ \bibnamefont {Thompson}},
  \bibinfo {author} {\bibfnamefont {R.~J.}\ \bibnamefont {Brecha}}, \bibinfo
  {author} {\bibfnamefont {W.~D.}\ \bibnamefont {Lee}}, \ and\ \bibinfo
  {author} {\bibfnamefont {H.~J.}\ \bibnamefont {Kimble}},\ }\bibfield  {title}
  {\enquote {\bibinfo {title} {Optical bistability and photon statistics in
  cavity quantum electrodynamics},}\ }\href {\doibase
  10.1103/PhysRevLett.67.1727} {\bibfield  {journal} {\bibinfo  {journal}
  {Phys. Rev. Lett.}\ }\textbf {\bibinfo {volume} {67}},\ \bibinfo {pages}
  {1727--1730} (\bibinfo {year} {1991})}\BibitemShut {NoStop}%
\bibitem [{\citenamefont {Collot}\ \emph {et~al.}(1993)\citenamefont {Collot},
  \citenamefont {Lefèvre-Seguin}, \citenamefont {Brune}, \citenamefont
  {Raimond},\ and\ \citenamefont {Haroche}}]{Collot93}%
  \BibitemOpen
  \bibfield  {author} {\bibinfo {author} {\bibfnamefont {L.}~\bibnamefont
  {Collot}}, \bibinfo {author} {\bibfnamefont {V.}~\bibnamefont
  {Lefèvre-Seguin}}, \bibinfo {author} {\bibfnamefont {M.}~\bibnamefont
  {Brune}}, \bibinfo {author} {\bibfnamefont {J.~M.}\ \bibnamefont {Raimond}},
  \ and\ \bibinfo {author} {\bibfnamefont {S.}~\bibnamefont {Haroche}},\
  }\bibfield  {title} {\enquote {\bibinfo {title} {Very high-{Q}
  whispering-gallery mode resonances observed on fused silica microspheres},}\
  }\href {http://stacks.iop.org/0295-5075/23/i=5/a=005} {\bibfield  {journal}
  {\bibinfo  {journal} {EPL (Europhysics Letters)}\ }\textbf {\bibinfo {volume}
  {23}},\ \bibinfo {pages} {327} (\bibinfo {year} {1993})}\BibitemShut
  {NoStop}%
\bibitem [{\citenamefont {Almeida}\ and\ \citenamefont
  {Lipson}(2004)}]{Lipson04}%
  \BibitemOpen
  \bibfield  {author} {\bibinfo {author} {\bibfnamefont {Vilson~R.}\
  \bibnamefont {Almeida}}\ and\ \bibinfo {author} {\bibfnamefont {Michal}\
  \bibnamefont {Lipson}},\ }\bibfield  {title} {\enquote {\bibinfo {title}
  {Optical bistability on a silicon chip},}\ }\href {\doibase
  10.1364/OL.29.002387} {\bibfield  {journal} {\bibinfo  {journal} {Opt.
  Lett.}\ }\textbf {\bibinfo {volume} {29}},\ \bibinfo {pages} {2387--2389}
  (\bibinfo {year} {2004})}\BibitemShut {NoStop}%
\bibitem [{\citenamefont {Notomi}\ \emph {et~al.}(2005)\citenamefont {Notomi},
  \citenamefont {Shinya}, \citenamefont {Mitsugi}, \citenamefont {Kira},
  \citenamefont {Kuramochi},\ and\ \citenamefont {Tanabe}}]{Notomi05}%
  \BibitemOpen
  \bibfield  {author} {\bibinfo {author} {\bibfnamefont {Masaya}\ \bibnamefont
  {Notomi}}, \bibinfo {author} {\bibfnamefont {Akihiko}\ \bibnamefont
  {Shinya}}, \bibinfo {author} {\bibfnamefont {Satoshi}\ \bibnamefont
  {Mitsugi}}, \bibinfo {author} {\bibfnamefont {Goh}\ \bibnamefont {Kira}},
  \bibinfo {author} {\bibfnamefont {Eiichi}\ \bibnamefont {Kuramochi}}, \ and\
  \bibinfo {author} {\bibfnamefont {Takasumi}\ \bibnamefont {Tanabe}},\
  }\bibfield  {title} {\enquote {\bibinfo {title} {Optical bistable switching
  action of {S}i high-{Q} photonic-crystal nanocavities},}\ }\href {\doibase
  10.1364/OPEX.13.002678} {\bibfield  {journal} {\bibinfo  {journal} {Opt.
  Express}\ }\textbf {\bibinfo {volume} {13}},\ \bibinfo {pages} {2678--2687}
  (\bibinfo {year} {2005})}\BibitemShut {NoStop}%
\bibitem [{\citenamefont {Wurtz}\ \emph {et~al.}(2006)\citenamefont {Wurtz},
  \citenamefont {Pollard},\ and\ \citenamefont {Zayats}}]{Wurtz06}%
  \BibitemOpen
  \bibfield  {author} {\bibinfo {author} {\bibfnamefont {G.~A.}\ \bibnamefont
  {Wurtz}}, \bibinfo {author} {\bibfnamefont {R.}~\bibnamefont {Pollard}}, \
  and\ \bibinfo {author} {\bibfnamefont {A.~V.}\ \bibnamefont {Zayats}},\
  }\bibfield  {title} {\enquote {\bibinfo {title} {Optical bistability in
  nonlinear surface-plasmon polaritonic crystals},}\ }\href {\doibase
  10.1103/PhysRevLett.97.057402} {\bibfield  {journal} {\bibinfo  {journal}
  {Phys. Rev. Lett.}\ }\textbf {\bibinfo {volume} {97}},\ \bibinfo {pages}
  {057402} (\bibinfo {year} {2006})}\BibitemShut {NoStop}%
\bibitem [{\citenamefont {Boulier}\ \emph {et~al.}(2014)\citenamefont {Boulier}
  \emph {et~al.}}]{Boulier14}%
  \BibitemOpen
  \bibfield  {author} {\bibinfo {author} {\bibfnamefont {T.}~\bibnamefont
  {Boulier}} \emph {et~al.},\ }\bibfield  {title} {\enquote {\bibinfo {title}
  {Polariton-generated intensity squeezing in semiconductor micropillars},}\
  }\href
  {http://www.nature.com/ncomms/2014/140212/ncomms4260/full/ncomms4260.html?message-global=remove}
  {\bibfield  {journal} {\bibinfo  {journal} {Nature Commun.}\ }\textbf
  {\bibinfo {volume} {5}} (\bibinfo {year} {2014})}\BibitemShut {NoStop}%
\bibitem [{\citenamefont {Risken}\ \emph {et~al.}(1987)\citenamefont {Risken},
  \citenamefont {Savage}, \citenamefont {Haake},\ and\ \citenamefont
  {Walls}}]{Risken87}%
  \BibitemOpen
  \bibfield  {author} {\bibinfo {author} {\bibfnamefont {H.}~\bibnamefont
  {Risken}}, \bibinfo {author} {\bibfnamefont {C.}~\bibnamefont {Savage}},
  \bibinfo {author} {\bibfnamefont {F.}~\bibnamefont {Haake}}, \ and\ \bibinfo
  {author} {\bibfnamefont {D.~F.}\ \bibnamefont {Walls}},\ }\bibfield  {title}
  {\enquote {\bibinfo {title} {Quantum tunneling in dispersive optical
  bistability},}\ }\href {\doibase 10.1103/PhysRevA.35.1729} {\bibfield
  {journal} {\bibinfo  {journal} {Phys. Rev. A}\ }\textbf {\bibinfo {volume}
  {35}},\ \bibinfo {pages} {1729--1739} (\bibinfo {year} {1987})}\BibitemShut
  {NoStop}%
\bibitem [{\citenamefont {Vogel}\ and\ \citenamefont {Risken}(1988)}]{Vogel88}%
  \BibitemOpen
  \bibfield  {author} {\bibinfo {author} {\bibfnamefont {K.}~\bibnamefont
  {Vogel}}\ and\ \bibinfo {author} {\bibfnamefont {H.}~\bibnamefont {Risken}},\
  }\bibfield  {title} {\enquote {\bibinfo {title} {Quantum-tunneling rates and
  stationary solutions in dispersive optical bistability},}\ }\href {\doibase
  10.1103/PhysRevA.38.2409} {\bibfield  {journal} {\bibinfo  {journal} {Phys.
  Rev. A}\ }\textbf {\bibinfo {volume} {38}},\ \bibinfo {pages} {2409--2422}
  (\bibinfo {year} {1988})}\BibitemShut {NoStop}%
\bibitem [{\citenamefont {Dykman}(2012)}]{Dykman12}%
  \BibitemOpen
  \bibfield  {author} {\bibinfo {author} {\bibfnamefont {Mark}\ \bibnamefont
  {Dykman}},\ }\href@noop {} {\emph {\bibinfo {title} {Fluctuating nonlinear
  oscillators: from nanomechanics to quantum superconducting circuits}}}\
  (\bibinfo  {publisher} {OUP Oxford},\ \bibinfo {year} {2012})\BibitemShut
  {NoStop}%
\bibitem [{sup()}]{supp}%
  \BibitemOpen
  \href@noop {} {}\bibinfo {note} {See Supplemental Material for details about
  the sample, setup characterization, noise measurements, mean-field
  calculations, scaling analysis, estimates of the polariton-polariton
  interaction constant in different micropillars, and comparison between
  measurements and calculations.}\BibitemShut {Stop}%
\bibitem [{\citenamefont {Dziarmaga}(2010)}]{Jacek10}%
  \BibitemOpen
  \bibfield  {author} {\bibinfo {author} {\bibfnamefont {Jacek}\ \bibnamefont
  {Dziarmaga}},\ }\bibfield  {title} {\enquote {\bibinfo {title} {Dynamics of a
  quantum phase transition and relaxation to a steady state},}\ }\href
  {\doibase 10.1080/00018732.2010.514702} {\bibfield  {journal} {\bibinfo
  {journal} {Adv. Phys.}\ }\textbf {\bibinfo {volume} {59}},\ \bibinfo {pages}
  {1063--1189} (\bibinfo {year} {2010})}\BibitemShut {NoStop}%
\bibitem [{\citenamefont {Jacqmin}\ \emph {et~al.}(2014)\citenamefont
  {Jacqmin}, \citenamefont {Carusotto}, \citenamefont {Sagnes}, \citenamefont
  {Abbarchi}, \citenamefont {Solnyshkov}, \citenamefont {Malpuech},
  \citenamefont {Galopin}, \citenamefont {Lema\^{\i}tre}, \citenamefont
  {Bloch},\ and\ \citenamefont {Amo}}]{Jacqmin}%
  \BibitemOpen
  \bibfield  {author} {\bibinfo {author} {\bibfnamefont {T.}~\bibnamefont
  {Jacqmin}}, \bibinfo {author} {\bibfnamefont {I.}~\bibnamefont {Carusotto}},
  \bibinfo {author} {\bibfnamefont {I.}~\bibnamefont {Sagnes}}, \bibinfo
  {author} {\bibfnamefont {M.}~\bibnamefont {Abbarchi}}, \bibinfo {author}
  {\bibfnamefont {D.~D.}\ \bibnamefont {Solnyshkov}}, \bibinfo {author}
  {\bibfnamefont {G.}~\bibnamefont {Malpuech}}, \bibinfo {author}
  {\bibfnamefont {E.}~\bibnamefont {Galopin}}, \bibinfo {author} {\bibfnamefont
  {A.}~\bibnamefont {Lema\^{\i}tre}}, \bibinfo {author} {\bibfnamefont
  {J.}~\bibnamefont {Bloch}}, \ and\ \bibinfo {author} {\bibfnamefont
  {A.}~\bibnamefont {Amo}},\ }\bibfield  {title} {\enquote {\bibinfo {title}
  {Direct observation of {D}irac cones and a flatband in a honeycomb lattice
  for polaritons},}\ }\href {\doibase 10.1103/PhysRevLett.112.116402}
  {\bibfield  {journal} {\bibinfo  {journal} {Phys. Rev. Lett.}\ }\textbf
  {\bibinfo {volume} {112}},\ \bibinfo {pages} {116402} (\bibinfo {year}
  {2014})}\BibitemShut {NoStop}%
\bibitem [{\citenamefont {Sala}\ \emph {et~al.}(2015)\citenamefont {Sala},
  \citenamefont {Solnyshkov}, \citenamefont {Carusotto}, \citenamefont
  {Jacqmin}, \citenamefont {Lema\^{\i}tre}, \citenamefont {Tercas},
  \citenamefont {Nalitov}, \citenamefont {Abbarchi}, \citenamefont {Galopin},
  \citenamefont {Sagnes}, \citenamefont {Bloch}, \citenamefont {Malpuech},\
  and\ \citenamefont {Amo}}]{Sala}%
  \BibitemOpen
  \bibfield  {author} {\bibinfo {author} {\bibfnamefont {V.~G.}\ \bibnamefont
  {Sala}}, \bibinfo {author} {\bibfnamefont {D.~D.}\ \bibnamefont
  {Solnyshkov}}, \bibinfo {author} {\bibfnamefont {I.}~\bibnamefont
  {Carusotto}}, \bibinfo {author} {\bibfnamefont {T.}~\bibnamefont {Jacqmin}},
  \bibinfo {author} {\bibfnamefont {A.}~\bibnamefont {Lema\^{\i}tre}}, \bibinfo
  {author} {\bibfnamefont {H.}~\bibnamefont {Tercas}}, \bibinfo {author}
  {\bibfnamefont {A.}~\bibnamefont {Nalitov}}, \bibinfo {author} {\bibfnamefont
  {M.}~\bibnamefont {Abbarchi}}, \bibinfo {author} {\bibfnamefont
  {E.}~\bibnamefont {Galopin}}, \bibinfo {author} {\bibfnamefont
  {I.}~\bibnamefont {Sagnes}}, \bibinfo {author} {\bibfnamefont
  {J.}~\bibnamefont {Bloch}}, \bibinfo {author} {\bibfnamefont
  {G.}~\bibnamefont {Malpuech}}, \ and\ \bibinfo {author} {\bibfnamefont
  {A.}~\bibnamefont {Amo}},\ }\bibfield  {title} {\enquote {\bibinfo {title}
  {Spin-orbit coupling for photons and polaritons in microstructures},}\ }\href
  {\doibase 10.1103/PhysRevX.5.011034} {\bibfield  {journal} {\bibinfo
  {journal} {Phys. Rev. X}\ }\textbf {\bibinfo {volume} {5}},\ \bibinfo {pages}
  {011034} (\bibinfo {year} {2015})}\BibitemShut {NoStop}%
\bibitem [{\citenamefont {Tanese}\ \emph {et~al.}(2014)\citenamefont {Tanese},
  \citenamefont {Gurevich}, \citenamefont {Baboux}, \citenamefont {Jacqmin},
  \citenamefont {Lema\^{\i}tre}, \citenamefont {Galopin}, \citenamefont
  {Sagnes}, \citenamefont {Amo}, \citenamefont {Bloch},\ and\ \citenamefont
  {Akkermans}}]{Tanese14}%
  \BibitemOpen
  \bibfield  {author} {\bibinfo {author} {\bibfnamefont {D.}~\bibnamefont
  {Tanese}}, \bibinfo {author} {\bibfnamefont {E.}~\bibnamefont {Gurevich}},
  \bibinfo {author} {\bibfnamefont {F.}~\bibnamefont {Baboux}}, \bibinfo
  {author} {\bibfnamefont {T.}~\bibnamefont {Jacqmin}}, \bibinfo {author}
  {\bibfnamefont {A.}~\bibnamefont {Lema\^{\i}tre}}, \bibinfo {author}
  {\bibfnamefont {E.}~\bibnamefont {Galopin}}, \bibinfo {author} {\bibfnamefont
  {I.}~\bibnamefont {Sagnes}}, \bibinfo {author} {\bibfnamefont
  {A.}~\bibnamefont {Amo}}, \bibinfo {author} {\bibfnamefont {J.}~\bibnamefont
  {Bloch}}, \ and\ \bibinfo {author} {\bibfnamefont {E.}~\bibnamefont
  {Akkermans}},\ }\bibfield  {title} {\enquote {\bibinfo {title} {Fractal
  energy spectrum of a polariton gas in a fibonacci quasiperiodic potential},}\
  }\href {\doibase 10.1103/PhysRevLett.112.146404} {\bibfield  {journal}
  {\bibinfo  {journal} {Phys. Rev. Lett.}\ }\textbf {\bibinfo {volume} {112}},\
  \bibinfo {pages} {146404} (\bibinfo {year} {2014})}\BibitemShut {NoStop}%
\bibitem [{\citenamefont {Akkermans}\ \emph {et~al.}(2009)\citenamefont
  {Akkermans}, \citenamefont {Dunne},\ and\ \citenamefont
  {Teplyaev}}]{Akkermans09}%
  \BibitemOpen
  \bibfield  {author} {\bibinfo {author} {\bibfnamefont {E.}~\bibnamefont
  {Akkermans}}, \bibinfo {author} {\bibfnamefont {G.~V.}\ \bibnamefont
  {Dunne}}, \ and\ \bibinfo {author} {\bibfnamefont {A.}~\bibnamefont
  {Teplyaev}},\ }\bibfield  {title} {\enquote {\bibinfo {title} {Physical
  consequences of complex dimensions of fractals},}\ }\href
  {http://stacks.iop.org/0295-5075/88/i=4/a=40007} {\bibfield  {journal}
  {\bibinfo  {journal} {EPL (Europhysics Letters)}\ }\textbf {\bibinfo {volume}
  {88}},\ \bibinfo {pages} {40007} (\bibinfo {year} {2009})}\BibitemShut
  {NoStop}%
\bibitem [{\citenamefont {Akkermans}\ \emph {et~al.}(2010)\citenamefont
  {Akkermans}, \citenamefont {Dunne},\ and\ \citenamefont
  {Teplyaev}}]{Akkermans10}%
  \BibitemOpen
  \bibfield  {author} {\bibinfo {author} {\bibfnamefont {Eric}\ \bibnamefont
  {Akkermans}}, \bibinfo {author} {\bibfnamefont {Gerald~V.}\ \bibnamefont
  {Dunne}}, \ and\ \bibinfo {author} {\bibfnamefont {Alexander}\ \bibnamefont
  {Teplyaev}},\ }\bibfield  {title} {\enquote {\bibinfo {title} {Thermodynamics
  of photons on fractals},}\ }\href {\doibase 10.1103/PhysRevLett.105.230407}
  {\bibfield  {journal} {\bibinfo  {journal} {Phys. Rev. Lett.}\ }\textbf
  {\bibinfo {volume} {105}},\ \bibinfo {pages} {230407} (\bibinfo {year}
  {2010})}\BibitemShut {NoStop}%
\end{thebibliography}

\begin{thebibliography}{6}%
\makeatletter
\providecommand \@ifxundefined [1]{%
 \@ifx{#1\undefined}
}%
\providecommand \@ifnum [1]{%
 \ifnum #1\expandafter \@firstoftwo
 \else \expandafter \@secondoftwo
 \fi
}%
\providecommand \@ifx [1]{%
 \ifx #1\expandafter \@firstoftwo
 \else \expandafter \@secondoftwo
 \fi
}%
\providecommand \natexlab [1]{#1}%
\providecommand \enquote  [1]{``#1''}%
\providecommand \bibnamefont  [1]{#1}%
\providecommand \bibfnamefont [1]{#1}%
\providecommand \citenamefont [1]{#1}%
\providecommand \href@noop [0]{\@secondoftwo}%
\providecommand \href [0]{\begingroup \@sanitize@url \@href}%
\providecommand \@href[1]{\@@startlink{#1}\@@href}%
\providecommand \@@href[1]{\endgroup#1\@@endlink}%
\providecommand \@sanitize@url [0]{\catcode `\\12\catcode `\$12\catcode
  `\&12\catcode `\#12\catcode `\^12\catcode `\_12\catcode `\%12\relax}%
\providecommand \@@startlink[1]{}%
\providecommand \@@endlink[0]{}%
\providecommand \url  [0]{\begingroup\@sanitize@url \@url }%
\providecommand \@url [1]{\endgroup\@href {#1}{\urlprefix }}%
\providecommand \urlprefix  [0]{URL }%
\providecommand \Eprint [0]{\href }%
\providecommand \doibase [0]{http://dx.doi.org/}%
\providecommand \selectlanguage [0]{\@gobble}%
\providecommand \bibinfo  [0]{\@secondoftwo}%
\providecommand \bibfield  [0]{\@secondoftwo}%
\providecommand \translation [1]{[#1]}%
\providecommand \BibitemOpen [0]{}%
\providecommand \bibitemStop [0]{}%
\providecommand \bibitemNoStop [0]{.\EOS\space}%
\providecommand \EOS [0]{\spacefactor3000\relax}%
\providecommand \BibitemShut  [1]{\csname bibitem#1\endcsname}%
\let\auto@bib@innerbib\@empty
\bibitem [{\citenamefont {Rodriguez}\ \emph {et~al.}(2016)\citenamefont
  {Rodriguez}, \citenamefont {Amo}, \citenamefont {Sagnes}, \citenamefont
  {Le~Gratiet}, \citenamefont {Galopin}, \citenamefont {Lemaitre},\ and\
  \citenamefont {Bloch}}]{Rodriguez16}%
  \BibitemOpen
  \bibfield  {author} {\bibinfo {author} {\bibfnamefont {S.}~\bibnamefont
  {Rodriguez}}, \bibinfo {author} {\bibfnamefont {A.}~\bibnamefont {Amo}},
  \bibinfo {author} {\bibfnamefont {I.}~\bibnamefont {Sagnes}}, \bibinfo
  {author} {\bibfnamefont {L.}~\bibnamefont {Le~Gratiet}}, \bibinfo {author}
  {\bibfnamefont {E.}~\bibnamefont {Galopin}}, \bibinfo {author} {\bibfnamefont
  {A.}~\bibnamefont {Lemaitre}}, \ and\ \bibinfo {author} {\bibfnamefont
  {J.}~\bibnamefont {Bloch}},\ }\href
  {http://www.nature.com/ncomms/2016/160616/ncomms11887/full/ncomms11887.html}
  {\bibfield  {journal} {\bibinfo  {journal} {Nature Commun.}\ }\textbf
  {\bibinfo {volume} {7}},\ \bibinfo {pages} {11887} (\bibinfo {year}
  {2016})}\BibitemShut {NoStop}%
\bibitem [{\citenamefont {Ciuti}\ \emph {et~al.}(1998)\citenamefont {Ciuti},
  \citenamefont {Savona}, \citenamefont {Piermarocchi}, \citenamefont
  {Quattropani},\ and\ \citenamefont {Schwendimann}}]{Ciuti98}%
  \BibitemOpen
  \bibfield  {author} {\bibinfo {author} {\bibfnamefont {C.}~\bibnamefont
  {Ciuti}}, \bibinfo {author} {\bibfnamefont {V.}~\bibnamefont {Savona}},
  \bibinfo {author} {\bibfnamefont {C.}~\bibnamefont {Piermarocchi}}, \bibinfo
  {author} {\bibfnamefont {A.}~\bibnamefont {Quattropani}}, \ and\ \bibinfo
  {author} {\bibfnamefont {P.}~\bibnamefont {Schwendimann}},\ }\href {\doibase
  10.1103/PhysRevB.58.7926} {\bibfield  {journal} {\bibinfo  {journal} {Phys.
  Rev. B}\ }\textbf {\bibinfo {volume} {58}},\ \bibinfo {pages} {7926}
  (\bibinfo {year} {1998})}\BibitemShut {NoStop}%
\bibitem [{\citenamefont {Abbaspour}\ \emph {et~al.}(2015)\citenamefont
  {Abbaspour}, \citenamefont {Sallen}, \citenamefont {Trebaol}, \citenamefont
  {Morier-Genoud}, \citenamefont {Portella-Oberli},\ and\ \citenamefont
  {Deveaud}}]{Deveaud15}%
  \BibitemOpen
  \bibfield  {author} {\bibinfo {author} {\bibfnamefont {H.}~\bibnamefont
  {Abbaspour}}, \bibinfo {author} {\bibfnamefont {G.}~\bibnamefont {Sallen}},
  \bibinfo {author} {\bibfnamefont {S.}~\bibnamefont {Trebaol}}, \bibinfo
  {author} {\bibfnamefont {F.}~\bibnamefont {Morier-Genoud}}, \bibinfo {author}
  {\bibfnamefont {M.~T.}\ \bibnamefont {Portella-Oberli}}, \ and\ \bibinfo
  {author} {\bibfnamefont {B.}~\bibnamefont {Deveaud}},\ }\href {\doibase
  10.1103/PhysRevB.92.165303} {\bibfield  {journal} {\bibinfo  {journal} {Phys.
  Rev. B}\ }\textbf {\bibinfo {volume} {92}},\ \bibinfo {pages} {165303}
  (\bibinfo {year} {2015})}\BibitemShut {NoStop}%
\bibitem [{\citenamefont {Kessler}\ \emph {et~al.}(2012)\citenamefont
  {Kessler}, \citenamefont {Giedke}, \citenamefont {Imamoglu}, \citenamefont
  {Yelin}, \citenamefont {Lukin},\ and\ \citenamefont {Cirac}}]{Kessler12}%
  \BibitemOpen
  \bibfield  {author} {\bibinfo {author} {\bibfnamefont {E.~M.}\ \bibnamefont
  {Kessler}}, \bibinfo {author} {\bibfnamefont {G.}~\bibnamefont {Giedke}},
  \bibinfo {author} {\bibfnamefont {A.}~\bibnamefont {Imamoglu}}, \bibinfo
  {author} {\bibfnamefont {S.~F.}\ \bibnamefont {Yelin}}, \bibinfo {author}
  {\bibfnamefont {M.~D.}\ \bibnamefont {Lukin}}, \ and\ \bibinfo {author}
  {\bibfnamefont {J.~I.}\ \bibnamefont {Cirac}},\ }\href {\doibase
  10.1103/PhysRevA.86.012116} {\bibfield  {journal} {\bibinfo  {journal} {Phys.
  Rev. A}\ }\textbf {\bibinfo {volume} {86}},\ \bibinfo {pages} {012116}
  (\bibinfo {year} {2012})}\BibitemShut {NoStop}%
\bibitem [{\citenamefont {Casteels}\ \emph {et~al.}(2016)\citenamefont
  {Casteels}, \citenamefont {Storme}, \citenamefont {Le~Boit\'e},\ and\
  \citenamefont {Ciuti}}]{Casteels16}%
  \BibitemOpen
  \bibfield  {author} {\bibinfo {author} {\bibfnamefont {W.}~\bibnamefont
  {Casteels}}, \bibinfo {author} {\bibfnamefont {F.}~\bibnamefont {Storme}},
  \bibinfo {author} {\bibfnamefont {A.}~\bibnamefont {Le~Boit\'e}}, \ and\
  \bibinfo {author} {\bibfnamefont {C.}~\bibnamefont {Ciuti}},\ }\href
  {\doibase 10.1103/PhysRevA.93.033824} {\bibfield  {journal} {\bibinfo
  {journal} {Phys. Rev. A}\ }\textbf {\bibinfo {volume} {93}},\ \bibinfo
  {pages} {033824} (\bibinfo {year} {2016})}\BibitemShut {NoStop}%
\bibitem [{\citenamefont {Luse}\ and\ \citenamefont {Zangwill}(1994)}]{Luse94}%
  \BibitemOpen
  \bibfield  {author} {\bibinfo {author} {\bibfnamefont {C.~N.}\ \bibnamefont
  {Luse}}\ and\ \bibinfo {author} {\bibfnamefont {A.}~\bibnamefont
  {Zangwill}},\ }\href {\doibase 10.1103/PhysRevE.50.224} {\bibfield  {journal}
  {\bibinfo  {journal} {Phys. Rev. E}\ }\textbf {\bibinfo {volume} {50}},\
  \bibinfo {pages} {224} (\bibinfo {year} {1994})}\BibitemShut {NoStop}%
\end{thebibliography}
\end{document}